\documentstyle[12pt,preprint,psfig]{aastex}

\newcommand{\kms}{km~s$^{-1}$}
\newcommand{\mup}{$M_{\rm{up}}$}
\newcommand{\mlow}{$M_{\rm{low}}$}
\newcommand{\ergsa}{erg~s$^{-1}$~\AA$^{-1}$}

\slugcomment{Submitted to The Astrophysical Journal, June 2002}

\lefthead{Gonz\'alez Delgado et al.}
\righthead{Seyferts 2}

\begin{document}

\title{The Massive Stellar Content in the Starburst NGC~3049:\\
 A Test for Hot-Star Models}

\author{Rosa M. Gonz\'alez Delgado}
\affil{Instituto de Astrof\'\i sica de Andaluc\'\i a (CSIC), Apdo. 3004, 18080 Granada, Spain}
\affil{Electronic mail: rosa@iaa.es}

\author{Claus Leitherer}
\affil{Space Telescope Science Institute, 3700 San Martin Drive, Baltimore, MD
21218}
\affil{Electronic mail: leitherer@stsci.edu}

\author{Gra\.{z}yna Stasi\'{n}ska}
\affil{LUTH, Observatoire de Paris-Meudon, 92195 Meudon Cedex, France}
\affil{Electronic mail: grazyna.stasinska@obspm.fr}

\and

\author{Timothy M. Heckman}
\affil{Department of Physics \& Astronomy, JHU, Baltimore, MD 21218}
\affil{Adjunct Astronomer at STScI}
\affil{Electronic mail: heckman@pha.jhu.edu}


\newpage

\begin{abstract}

The objective of this work is twofold: First, we seek 
evidence for or against the depletion of massive stars in metal-rich starbursts.
A second, equally important goal is to 
perform a consistency test of the latest generation of starburst models 
in such a high-metallicity environment.
We have obtained high-spatial resolution ultraviolet and optical STIS 
spectroscopy and imaging of the metal-rich nuclear starburst in NGC~3049.
The stellar continuum and the absorption line spectrum in the ultraviolet
are used to constrain the massive stellar population. The strong,
blueshifted stellar lines of CIV and SiIV detected in the UV spectra 
indicate a
metal-rich, compact, massive ($\sim$10$^6$~M$_\odot$) cluster of age 3~--~4~Myr 
emitting the UV-optical
continuum. We find strong evidence against a depletion of 
massive stars in 
this metal-rich cluster. The derived age and the upper mass-limit cut-off of 
the initial mass function are also
consistent with the detection of Wolf-Rayet (WR) features at optical wavelengths. As a second independent 
constraint on the massive stellar content, 
the nebular emission-line spectrum is modeled with photoionization codes 
using stellar spectra from evolutionary synthesis models. The morphology of the nuclear 
starburst of NGC~3049 from the STIS images indicates a simple geometry for
the nebular emission-line region.
However, the nebular lines are badly reproduced 
by 3~--~4~Myr instantaneous bursts, as required by the UV line spectrum, 
when unblanketed WR and/or Kurucz stellar atmospheres are used.
The corresponding number of photons above
24 and 54~eV in the synthetic models is too high in comparison with values suggested
by the observed line ratios.
Since the ionizing spectrum in this regime is dominated by emission from WR stars, this discrepancy
between observations and models is most likely the result of incorrect
assumptions about the WR stars. Thus we
conclude that the nebular 
spectrum of high-metallicity starbursts is poorly reproduced by models for
WR dominated populations. However, the new model set of Smith et al. (2002)
with blanketed WR and O atmospheres and adjusted WR temperatures
predicts a softer far-UV radiation field, providing a better match to the data.

\end{abstract}



\newpage

\section{Introduction}

Starbursts are the site where most of the high-mass star formation occurs. They appear
in disks, in bulges and in the nuclei of different types (e.g., spirals, 
irregulars, or dwarfs) of nearby 
and distant galaxies. Terlevich (1997) defined 
starburst galaxies as objects in which the energy output of the starburst dominates that 
of the host galaxy. Thus, starbursts play a major role in the processes of evolution
and formation of galaxies due to their high star-formation activity. They constitute ideal
laboratories to investigate some key issues: the formation and evolution of massive stars; 
the feedback between the interstellar medium and the star formation processes; or the star 
formation and chemical evolution of the universe. 

One important question is which stars form in starbursts. The
amount of mass transformed into stars and the mass range of the newly formed
stars determine the period of time over which a 
galaxy can support a starburst phase, and therefore the effect of starbursts
on the evolution of
galaxies. Thus, it is crucial to know if starbursts have an extreme initial mass function 
(IMF) with respect to more quiescent systems. 

Numerous studies performed in the last few years suggest that the IMF has an universal 
nature, having a slope 
close to Salpeter for a mass range between 5~M$_\odot$ and 60~M$_\odot$ (e.g., references in Gilmore 
\& Howell 1998).
However, the IMF in starbursts is still not well known, in particular in high-metallicity 
environments. Contradictory results have been presented.
An extreme IMF for starbursts was suggested by Rieke et al. (1980) in a pioneering work
on M82. They proposed a {\it top heavy} IMF (with a deficit of stars below 3~--~8~M$_\odot$)
to explain the mass-luminosity ratio estimated 
from K-band photometry and a dynamical mass measurement. 
This IMF includes a higher fraction of red supergiant stars over red giants and 
dwarfs than expected for a normal IMF. Their results have been confirmed  
by Rieke et al. (1993). Satyapal et al. (1997) and Foerster-Schreiber (2000) 
accounted for the K-band luminosity 
with a Salpeter IMF and ascribe this apparently discrepant result
to the complex morphology of M82 and to dust obscuration in the starburst.      

The analysis of the nebular 
optical+near-infrared (IR) lines indicates the suppression of stars more massive than 30~M$_\odot$
(Goldader et al. 1997; Bresolin, Kennicutt, \& Garnett 1999;  Coziol, Doyon, \& Demers 2001).
Thornley et al. (2000) used the mid-infrared line ratio of 
[NeIII]$\lambda$15.6/[NeII]$\lambda$12.8 to measure the hardness of the stellar ionizing 
radiation. They found that, on average, stars with masses above about 40~M$_\odot$ 
are not present in starburst galaxies, either because they were never formed
or because they already have disappeared as a result of aging effects. 
A complication of the interpretation of [NeIII]/[NeII] are the uncertain
stellar evolution models at high metallicity. As pointed out by
Thornley et al., observations of gas and stars in the Galactic Center suggest
a  disagreement between the tracks and the observations, which  may be related to 
the difficulty of defining mass loss, atmospheres, and effective temperatures 
for metal-rich stars.

On the other hand, the UV-optical continuum from the integrated spectrum of starbursts 
suggests the 
formation of such massive stars (Leitherer 1996; Schaerer 2000; Gonz\'alez Delgado 2001). This
technique has the advantage of probing the stellar light from massive O stars directly,
as opposed to a nebular analysis which relies on an indirect measure of the stellar
radiation. Using WR-star features,
Schaerer et al. (2000) find that a Salpeter IMF extending to masses \mup~$\geq$~40~M$_\odot$
is compatible with the WR-star census  observed in metal-rich starbursts.
A similar conclusion has been obtained by Bresolin \& Kennicutt (2002) 
and by Pindao et al. (2002) based on the detection of WR features 
in high-metallicity HII regions.

Most of the results on the IMF have been obtained from a limited number of constraints, 
through the 
modeling of a few starburst properties at a specific wavelength. These results can depend strongly on the 
model ingredients. In particular, the evolutionary synthesis models used 
to analyze the UV+optical+near-IR properties of the starbursts depend on the stellar tracks, atmospheres,
and libraries (see the discussion in Schaerer 2000). 
In addition, the modeling of the nebular lines depends on the assumptions of the gas geometry, 
the electron 
density structure, and the chemical composition of the gas. Thus, inclusion of as
many different starburst 
properties as possible is required for a determination of the IMF.

We have obtained UV and optical spectra and images of the starburst galaxy NGC~3049 
with Space Telescope Imaging Spectrograph (STIS) on board of the Hubble Space Telescope (HST)
to investigate the possible evidence of depletion of massive stars in metal-rich starbursts.
  NGC~3049 is a barred spiral galaxy, SB(rs)ab, in the Virgo cluster, known as Mrk~710 in the 
Markarian \& Lipovetskii (1976) catalog. It is also classified as a nuclear starburst (Balzano 1983),
a WR galaxy (Conti 1991) and an HII galaxy (Terlevich et al. 1991). 
Kunth \& Schild (1986) first reported the detection
of broad emission features at NIII $\lambda$4640 and HeII $\lambda$4686 produced by WR stars.
More recently, Schaerer, Contini, \& Kunth (1999) reported the detection of broad CIV $\lambda$5808
indicating the presence of WC in addition to late WN stars
 in the starburst. Due to its nebular emission lines and WR
features, NGC~3049 has been observed intensively in the past (e.g. Masegosa, Moles, \& del Olmo 1991;
Vacca \& Conti 1992; Storchi-Bergmann et al. 1994, 1995; Contini 1996; Guseva, Izotov, \& Thuan 2000). 
NGC~3049 is  very bright in the far-IR ($L_{\rm{IR}} = 9 \times 10^{42}$~erg~s$^{-1}$; Heckman 
et al. 1998) and at UV wavelengths (Kinney et al. 1993). Its IUE spectrum
suggests a young stellar population dominating the central (20$\times$10 arcsec) emission
(Mas-Hesse \& Kunth 1999). Optical images show extended recent star formation  along the bar
(Mazzarella \& Boronson 1993; Contini et al. 1997; Schaerer et al. 1999). 
Molecular gas has been detected at the center of the galaxy, but no dense gas as traced by the HCN 
or CS molecules was found (Contini et al. 1997).
The oxygen abundance of the nucleus has been estimated by the 
{\it strong line} semi-empirical methods because auroral lines (e.g., [OIII] $\lambda$4363, [NII] $\lambda$5755, 
[SIII] $\lambda$6312) that are used as electron temperature diagnostics have not been detected in the spectrum.
These methods rely, e.g., on $R_{23}$=([OII]$\lambda$3727+[OIII]$\lambda$5007)/H$\beta$ or 
[NII]$\lambda$6584/H$\alpha$ to derive a supersolar oxygen abundance. 
Vacca \& Conti (1992) give
$12+\log$(O/H)~=~9.08, Storchi-Bergmann et al. (1994) give 8.87, Guseva et al. (2000) give 9.03, and Contini
(1996) gives 9.05\footnote{These abundances have been estimated using the $R_{23}$ calibration of Edmund \& Pagel 
(1984) or the relation between the oxygen abundance and the [NII]$\lambda$6584/H$\alpha$ ratio derived by 
van Zee et al (1998), that was obtained also using the Edmund \& Pagel calibration}.
Even considering the uncertainties associated with these methods, we are confident that
the metallicity in the nucleus of NGC~3049 is supersolar ($12+\log$(O/H)~$ \geq 8.9$).

NGC~3049 has been chosen for this project  because previous WFPC (pre-Costar) images at UV wavelengths
indicate a relatively simple morphology, with the UV flux dominated by one extremely bright 
star cluster. This morphology minimizes geometry effect and it makes it easier to study the relation between the
stellar and the nebular spectrum, and to search for evidence of an extreme IMF in NGC~3049. In addition, our goal is 
to perform a critical test of the most sophisticated hot-star models currently available to 
predict the UV stellar properties and the optical nebular lines in metal-rich starbursts. Our work is
organized as follows: the observations and data reduction are in Section~2; in Section~3, we describe the 
morphology of the UV and optical emission; Sections~4 to 6 deal
with the analysis and interpretation of the UV light, UV-optical continuum, and nebular optical lines, respectively.
The summary and conclusions are in Section~7.

\section{Observations and data reduction}

We obtained UV and optical spectra and images of NGC~3049 with STIS on board the HST during two visits.
In the first visit, on 1998 November 4, NGC~3049 was imaged in the far-UV with the FUV-MAMA detector through
the filter F25SRF2 ($\lambda_{\rm c}=1469$~\AA), and at optical wavelengths with the CCD detector through the filters 
F28$\times$50LP ($\lambda_{\rm c}=7150$~\AA) and F28$\times$50OIII ($\lambda_{\rm c}=$5007~\AA).
The total integration time was 900~s in the far-UV. In the optical, the total integration time of
240~s (with the optical longpass filter) and 948~s (with the [OIII] filter) was split in 4 and 2 individual
exposures, respectively. The spatial sampling of the FUV-MAMA and CCD detectors is 0.0244 arcsec/pixel 
and 0.05 arcsec/pixel, respectively. The individual images were calibrated with the standard STScI pipeline,
and they were combined with the cosmic-reject command to generate an individual image in the UV and optical  
(mainly H$\alpha$ plus red continuum), and at [OIII] plus the underlying continuum. 

In the second visit, on 1999 January 18, UV and optical spectra of the central region of NGC~3049 were taken.
We used the STIS/FUV-MAMA detector with a 52$\times$0.5 arcsec slit and the G140L grating. 
A total integration time of 11064~s was split in four individual exposures. The spectra cover the wavelength
range of 1150~--~1700~\AA\ with a dispersion of 0.6~\AA/pixel, giving a minimum spectral resolution of $\sim$1~\AA\ 
(for point sources, and larger, up to $\sim$12~\AA\, for extended sources).
At optical wavelengths, the STIS/CCD detector was used with a 52$\times$0.1 arcsec slit. Note, 
however, that the different slit widths used for the optical and UV spectroscopy  pose no problem in the
analysis of the spectral energy distribution (UV-optical continuum), as we demonstrate in section 5.  An
exposure with a total integration
time of 810~s was taken with the G430L grating covering the wavelength range of 2900~--~5700~\AA\ with a dispersion 
of 2.7~\AA/pixel. This time was split in two individual exposures. In addition, we took a 972~s 
exposure with the 
G750M grating centered on 6581~\AA\ and covering 6300 to 6870~\AA\ with a dispersion of 0.6~\AA/pixel. 
This exposure was
split in two individual sub-exposures as well. 
The spectral resolution of the optical spectra is 4~--~5.4~\AA\ and 0.8~--~1.1~\AA\ in the blue and 
red, respectively.  The slit was centered on the brightest nuclear knot and oriented along the bar
at P.A.~=~$-138^{\circ}$. The data extend over more than 8~arcsec with a scale of 0.0244~arcsec/pixel 
at the UV continuum, and 11~arcsec with a scale of 0.05~arcsec/pixel at the H$\alpha$ line.

The individual spectra were first calibrated with the standard STScI pipeline. Then, the calibrated frames 
were combined to produce a single two-dimensional spectrum at UV, blue, and red optical wavelengths.  
 We have subtracted 
a one-dimensional spectrum obtained from the outer part of the UV 2D frame
to correct the UV spectrum for the strong geocoronal emission at Ly$\alpha$ and OI $\lambda$1302.
This spectrum has only a 
very weak continuum
plus the Ly$\alpha$  and OI and is a good approximation for a sky spectrum.  
All the spectra have been treated as extended continuum 
sources when converting from surface brightness units in the 2D frames to flux units 
(erg~s$^{-1}$~cm$^{-2}$~\AA$^{-1}$).
Finally, the spectra have been de-redshifted assuming a heliocentric radial velocity of 1494~km~s$^{-1}$.

In addition, we have retrieved narrowband H$\alpha$ and nearby continuum images from the data archive of the 
Isaac Newton Group of Telescopes (ING, La Palma). The galaxy was observed on 2001 January 1 with 
the 1.0~m Jacobus Kapteyn
Telescope (JKT) using a 2088$\times$2120 SIT CCD as a detector, with a spatial sampling of 0.33~arcsec/pixel, 
through an H$\alpha$ ($\lambda_{\rm c}= 6594$~\AA, FWHM~=~44~\AA) and a nearby 
continuum filter ($\lambda_{\rm c}= 6656$~\AA, 
FWHM~=~44~\AA). The integration time was 2400~s in each filter, split in two exposures. The nearby continuum
was subtracted iteratively from the H$\alpha$+continuum image to produce a pure H$\alpha$ image. (Note, however, that this
image contains also part of the [NII] emission). The final image is displayed in Figure~1.

\section{The structure of NGC 3049}

\subsection{Morphology at the UV and optical wavelengths}

The JKT H$\alpha$ image reveals that most of the emission is along the bar and some arcs that emerge from the ends of the bar
with a ring-like appearance of 30~arcsec radius (Figure~1). The highest H$\alpha$ surface brightness is at the center 
of the stellar bar, the nucleus of the galaxy, and at 10~arcsec north-east.   

The central 20$\times$20 arcsec far-UV and optical HST images of NGC~3049 are in Figure~2. Both images
show extended emission along the bar, with a similar appearance. However, a clear difference between the two images 
is the emission at $\sim$2~arcsec NE, which is brighter at optical than at UV wavelengths. This could be due
to an older stellar population contributing to the red-optical continuum or due to higher obscuration in the 
NE dimming the UV emission. 

The magnitude  in the HST system measured in a circular aperture of 0.25~arcsec radius (10~arcsec) 
is 14.67 and 17.60 (13.40 and 14.45) at UV and 
optical wavelengths, respectively. Table~1 gives the UV and optical magnitudes  of NGC~3049
 as a function of the
distance obtained from the circular aperture photometry in the UV and optical images.

An expansion of the central 4$\times$4~arcsec of NGC~3049 is shown in Figure~3. The [OIII] 
plus the underlying continuum image is reproduced in this figure as well.
The three images suggest that the nuclear starburst is compact, and most of the emission is dominated
by a very bright central cluster. However, at least 6 clusters, much weaker, are also detected in the center of the galaxy.
The observed properties of these clusters  
are in Table~2. (The location and precise extraction areas are shown in Figures~3 and 4).
The similarity between the UV and optical central morphology indicates
that there is not much obscuration at the nucleus of NGC~3049. It looks like hot stars and ionizing gas 
can be described by a simple geometry. Thus, we expect that the effect of the geometry
on the modeling of the nebular lines should be small.

\subsection{The UV and optical spectra}

Figure~4 shows the integrated UV flux as a function of the slit position. The UV flux is 
dominated by the central cluster, but there are other sources embedded in the extended diffuse emission within 
the central $\sim$2.5~arcsec, as clearly shown in the UV image (Figure~3).  
As expected from the morphology
of the images, the integrated optical continuum (near H$\alpha$) and  the H$\alpha$ emission have
a spatial distribution similar to the UV flux.

As a compromise between signal-to-noise and spectral resolution, we have obtained the UV spectra 
with a slit width of 0.5~arcsec, larger than the 0.1~arcsec width used in the optical spectra. To check
the effect of the different slit widths used, we have extracted the spatial profile along 
P.A.~=~$-138^{\circ}$ in the UV 
and optical images, simulating a slit of width 0.1 and 0.5~arcsec. Both profiles follow the same spatial distribution,
and the spectrum of the central 0.3~arcsec is hardly affected by the slit width 
mismatch because the central 
cluster dominating the emission is very compact. The FWHM of the spatial profile corresponding to the 
central knot is only 0.14 and 0.12~arcsec in the optical and UV images, respectively. Of course, 
outside the central
cluster, the UV and optical slits are not sampling the same spatial distribution. 

Several one-dimensional spectra have been extracted. The windows used at the UV wavelengths are marked in Figure~4. 
The corresponding one-dimensional red and blue optical spectra (a and b) have been also extracted. Figures~5
 and 6 show
these spectra. The UV spectrum is dominated by absorption lines formed in the interstellar medium, in the winds,
 and in the photosphere 
of massive stars. The main interstellar lines detected in the nuclear spectrum of NGC~3049 are 
SiII $\lambda$1260, OI+SiII $\lambda$1302, CII $\lambda$1335, SiII $\lambda$1526, FeII $\lambda$1608, and 
AlII $\lambda$1670; the wind lines are NV $\lambda$1240, SiIV $\lambda$1400, and CIV $\lambda$1550,( 
but HeII $\lambda$1640 is not clearly detected);
and the photospheric lines are OV $\lambda$1371, FeV $\lambda$1360--1380, SiIII $\lambda$1417, 
CIII $\lambda$1427, and SV $\lambda$1501. The strength of the photospheric lines and the P~Cygni profile of the
wind lines indicates that a young powerful starburst dominates the UV light. The optical spectra show a strong
blue continuum provided by young (O and B) stars, and broad emission lines at $\sim\lambda$4660 \AA\ from 
WR stars. No absorption lines from an older stellar population have been detected; thus,
the optical continuum  is only produced by the very young starburst as seen at the UV.
 The contribution from the bulge
population is negligible through this narrow slit. 
The optical spectra show nebular lines from the Balmer series such as
H$\alpha$, H$\beta$ and H$\gamma$, low ionization lines such as [OII] $\lambda$3727, [NII] $\lambda$6584,6543,
[SII] $\lambda$6717,6731, and the high-excitation line [OIII]$\lambda$5007.

\subsection{WR-star distribution}

As expected, broad emission lines due to WR stars are detected in the STIS blue optical spectrum
at $\sim$4660~\AA. The STIS red spectrum does not cover the range 5700~--~6300~\AA. Therefore
we cannot confirm the detection of 
the red WR bump at 5810~\AA, which was
previously detected from ground-based observations (Schaerer et al. 1999; Guseva et al. 2000).
The blue bump is mainly formed by the blend of NIII $\lambda$4620--4640, CIII 
$\lambda$4650+CIV $\lambda$4658 and 
HeII $\lambda$4686 (Figure~6). NV $\lambda$4604 or NIV $\lambda$4507 are not detected, but a weak SiIII 
$\lambda$4552--4576 
blend may be present. This implies that late WN and early and late WC stars are the dominant WR stellar population. However, the weak
SiIII blend indicates that some early WN stars can also be present. The detection of 
CIII $\lambda$4650, $\lambda$5696 (Schaerer et al. 1999) 
suggests a supersolar metallicity of the starburst because late-type WC stars are only 
seen in high-metallicity
regions (Smith et al. 1991; Phillips \& Conti 1992).  

No {\em nebular} HeII $\lambda$4686 is detected in the spectra. 
However, considering that this line is 0.01~--~0.025 the intensity 
of H$\beta$ (Guseva et al. 2000), we could not detect it in our STIS spectra even if it were present.
We have not unambiguously detected {\em broad} 
HeII $\lambda$1640 in the STIS UV spectra, either. This line must be present, given the 
presence of the stellar $\lambda$4686 line. However, this region is at the edge of the
wavelength coverage of the far-UV spectrum, where the S/N is lower and where
the continuum placement is challenging.

We have measured the 4660~\AA\ bump in each pixel along the spatial direction to trace the spatial distribution of the 
WR stars. We find that the flux of the bump peaks at the maximum of the nearby optical continuum.
Consequently the WR stars have the
same spatial distribution as the main sequence stars. The bump is only detected in 6~pixels ($\sim$0.3~arcsec),
indicating a very compact spatial distribution of the WR stars. 

The total flux of the bump (between 4600 and 4720~\AA) is $1.1\times 10^{-14}$~erg~s$^{-1}$~cm$^{-2}$, and the 
equivalent width is $\sim$14~\AA. This flux is a factor of 2 lower than the flux measured in ground-based observations
(Schaerer et al. 1999) with an aperture of 2~arcsec. However, given the sharp spatial distribution of the WR bump
and the continuum, we do not think that differences in fluxes can be produced by a more extended WR stellar population.
The fluxes of the ground-based spectra have quoted uncertainties of about 30\%. Therefore we believe that the
discrepancy between the HST and ground-based data is the result of observational uncertainties. We note that 
the equivalent widths agree rather well, with the HST value being somewhat larger than the ground-based
value of 11~\AA.

\subsection{Kinematics of the ionized gas}

We constructed the ionized gas velocity curve by fitting a Gaussian to the H$\alpha$ emission line
 detected along 
the stellar bar, out to $-4$~arcsec 
south-west and +11~arcsec north-east of the nucleus (defined as the peak of the optical
continuum). The ionized gas does not follow a rotation curve (Figure~7a), even though our 
observing angle is close to the position of the line of nodes, 
P.A.~=~25$^{\circ}$, and the disk inclination is 55$^{\circ}$. Under these conditions we would expect a
rotation curve close to the maximum amplitude.

The velocity of the gas close to the nucleus is 1580~\kms, $\sim$90~\kms\ larger than the systemic velocity given
in NED, 1494~\kms, and it is about 40~\kms\ lower than the velocity measured by Schaerer et al. (1999). The 
general large-scale shape is in good agreement with Schaerer et al. (1999). However, our data
 show more kinematic structure 
in the central 2~arcsec (Figure~7b). At 0.5~arcsec north-east of the nucleus, the gas 
is blueshifted 
by 100~\kms, 
and at 1~arcsec, the gas has again the nuclear velocity. Therefore,  the gas is both approaching
and receding along the line of sight by 100~\kms\ within a spatial scale of 1~arcsec.

Two kinematic components are resolved and extended by 0.4~arcsec south-west of the nucleus, and by 
0.1~arcsec
at 0.6~arcsec north-east. These two components are also detected in [NII] $\lambda$6584,6548 (Figure~8), 
but not in H$\beta$
or [OIII]. In the south-west, the second weaker component is blueshifted by $\sim$60~\kms\ with respect 
to the main brightest
component with a velocity of 1600~\kms. This component may represent a bubble powered
 by the winds from the massive stars.

Other evidence of the interaction of massive stars with the interstellar medium comes
from the UV interstellar lines.
Some of these lines are detected at zero redshift (in the observed spectrum {\it a} and {\it b}), and thus are associated 
with gas in the Milky Way. However, other lines, such as SiII $\lambda$1260, OI+SiII $\lambda$1303, CII $\lambda$1335, and 
SiII $\lambda$1526, form in the interstellar medium of the starburst. We have measured the radial velocity of   
 SiII $\lambda$1260 and CII $\lambda$1335 with respect to the Milky Way (MW) lines. Their radial velocities indicate
blueshifts of about 210~\kms\ with respect to the redshift given in NED and about 300~\kms\ with respect to
the radial velocity measured in the nebular ionized gas. Thus, the warm interstellar gas in the nuclear starburst of 
NGC~3049 is outflowing by 200~--~300~\kms. Outflows with similar velocities have been detected in other starburst
galaxies (e.g., Gonz\'alez Delgado et al. 1998; Kunth et al. 1998; Johnson et al. 2000).

\section{Modeling the UV light}

The absorption lines formed in the wind of massive stars are driven by radiation pressures; their profiles depend 
on the luminosity  of the star (Castor, Abbott, \& Klein 1975). Therefore, since there is a well
defined stellar mass-luminosity relation, the profiles in the integrated light of a starburst depend on the 
IMF, star formation history, and age of the stellar clusters (Leitherer, Robert, \& Heckman 1995). Thus, 
the profile shapes
of the main wind lines such as NV, SiIV, and CIV can be used to constrain the properties of the starburst. 

\subsection{Description of the models}

We compare the observed UV spectra (wind and photospheric lines,
and UV continuum flux) to the predictions from evolutionary stellar population models. We use the code STARBURST99
(hereafter SB99; Leitherer et al. 1999) which is optimized to reproduce many properties of active star forming galaxies.
The stellar population is parameterized in terms of an IMF which  follows a power-law between two
mass limits \mup\ and \mlow. The stellar population is evolved from the zero-age main sequence using the 
zero-rotation stellar tracks of the Geneva group (Meynet et al. 1994).
The SB99 models use two extreme star formation histories, an instantaneous burst and constant star formation (csf). 
 The code can use two  stellar libraries in this spectral region, depending on the metallicity. 
One of these libraries is built with IUE spectra of Milky Way stars
 in the solar vicinity with solar or slightly subsolar metallicity. The O and WR stellar spectra are from the atlas of 
Robert, Leitherer, \& Heckman (1993) and the
B spectra from de~Mello, Leitherer, \& Heckman (2000).  The second library has 
been recently built with an atlas of metal-poor ($Z\approx 1/4$~Z$_\odot$) O stars in the Large and Small Magellanic Clouds 
(Leitherer et al. 2001). 
Models are computed for instantaneous bursts between 0 and 10 Myr, and for csf lasting for 10 Myr, and 
for different assumptions
about the slope ($\alpha=2.35$, 1.5) and upper mass cut-off (\mup~=~100, 60, and 40~M$_\odot$) of the IMF.

\subsection{Model results}

As the strength of the P~Cygni features depends on the mass-loss rate, which is a function of the metallicity 
(Puls et al. 1996), we have computed models using the solar metallicity stellar tracks and  the MW stellar library, and 
the subsolar stellar tracks ($Z$~=0.008 and 0.004) and the Magellanic Clouds stellar library. This has been done to confirm that 
the starburst has supersolar metallicity instead subsolar.  Models with subsolar 
metallicity and using the MC stellar library 
predict weaker wind lines  than the lines detected in the spectrum of NGC~3049. Thus, the nuclear stellar metallicity 
of NGC 3049 is at least  solar, consistent with the result from the ionized gas. Therefore, the models are 
computed with the MW stellar library and the stellar evolutionary tracks
at $Z$~=~0.02 and $Z$~=~0.04 in order to constrain
the star formation history, age, and IMF.
   
We can use the relative strength of the 
SiIV P~Cygni profile with respect to the P~Cygni profiles of CIV and NV in order
to constrain the star-formation history of the youngest stellar population.
SiIV has a strong profile only in O supergiants,
while CIV and NV are strong in O stars of any luminosity class. 
Thus, SiIV has a strong P~Cygni profile if the cluster formed
in an instantaneous burst and its age is between 3 and 5 Myr. On the other hand, the profile of CIV depends 
strongly on the slope and upper mass limit of the IMF. 

SiIV is very strong in the nuclear (cluster {\it b}) spectrum. The intensity of the emission and the depth of the absorption 
of the profile are only  reproduced if many O blue supergiants dominate the UV light, requiring 
instantaneous burst models. In clusters {\it c} and {\it d}, the SiIV is weaker, indicating 
a lower  fraction of O  
supergiants
with respect to O main-sequence stars than in cluster {\it b}. However, the P~Cygni profile is also well fitted by instantaneous burst models.
Because cluster {\it b} dominates the total central UV light, and wind lines are not diluted in extraction 
{\it a} with 
respect to {\it b}, the best models to predict the wind lines in the central spectrum are also instantaneous bursts.
Therefore, the UV central light is provided by clusters of similar age.

We deduce clusters ages of 3~--~4~Myr from the comparison of the wind profiles with instantaneous burst models
if we assume that the stars evolved following the $Z$~=~0.02 or $Z$~=~0.04 stellar tracks. 
Models with a Salpeter IMF with \mup~$\geq 60$~M$_\odot$
fit well the wind lines; however, models with a flatter Salpeter IMF (and  with a larger fraction
 of massive stars)
are also compatible with the observed wind lines (Figure~9). However, because of
 the strong P~Cygni profile of SiIV and CIV, we can 
rule out clusters younger than 2.5~Myr and older than 4~Myr, and models with very few massive stars 
(e.g., IMF with \mup~=~40~M$_\odot$;
Figure~10). We conclude, that in spite of the high-metallicity environment in the
center of the galaxy, very massive stars formed in these clusters, and they formed with a very small age spread.  

 We can compare the reddening corrected UV continuum luminosity provided by the clusters 
with the predictions from evolutionary models for a mass estimate.
The UV intrinsic spectral distribution of young unobscured  clusters follows
a power law ($F_{\rm{UV}}\propto {\lambda}^{\beta}$), 
with $\beta$ depends little on  the IMF and the 
star-formation history. 
A deviation of the exponent from the predicted values can be interpreted as a dust effect. 
Note, however, that the UV light does not account for low-mass stars. Their contribution must be
added by assuming an IMF at the low-mass end. In addition,
our estimate
depends on the reddening correction applied to the observed UV flux, and therefore on the 
dust obscuration law.

Using the reddening law of Calzetti et al. (2000), which is an empirical attenuation
 curve derived for starbursts, we estimate 
$E(B-V)=0.2$ to match the observed spectra with the UV spectral energy distribution 
(SED) predicted by the models. Because the
Galactic foreground extinction is very small, $E(B-V)=0.01$, most of the reddening is intrinsic to the starburst.
There is no significant change of the reddening in the center of NGC~3049, as suggested
by the same value for the different UV spectra extracted ({\it a, b, c}, and {\it d}). 

After correcting for reddening and assuming a distance of 18~Mpc 
(Vacca \& Conti 1992), the UV luminosity at 1500~\AA\ is $\log L_{1500}$= 39.52 (erg~s$^{-1}$~\AA$^{-1}$) 
and 39.35 (\ergsa) for the spectra {\it a} and {\it b}, 
respectively. Therefore, the mass of the central starburst is $\sim$10$^6$~M$_\odot$, and the number of O stars 
in the central cluster is $\sim$4000. 
This mass is only a factor of two lower than the stellar mass required to 
account for the total bolometric luminosity (assuming $\log L_{\rm{Bol}} = \log L_{\rm{FIR}}=
42.93$ (erg~s$^{-1}$)). This cluster, together with the other fainter clusters detected in the
UV, can come up for almost half of the IRAS luminosity of NGC~3049, suggesting that a significant fraction of  the UV
is reprocessed into far-IR luminosity. The stars seen via their spectral lines in the UV 
are representative of the global starburst. There is no need to hypothesize an additional,
dust obscured population. 
We can estimate the number of WR stars and the WR/O ratio from the WR blue bump luminosity.
The luminosity of the bump after correcting for reddening ($E(B-V)=0.2$) is $\log L_{4660}= 38.94$ (erg~s$^{-1}$).
This translates into $\sim$275
WR stars in the cluster if we assume that the average luminosity of a WN7 star
 is $\log L_{4660}= 36.5$ (erg~s$^{-1}$).
The resulting  WR/O ratio of 0.065 is in agreement with the prediction by an instantaneous 
burst of age 3~Myr (WR/O~=~0.08 for $Z$~=~0.02).
We also estimated the number total ionizing photons (see Table~2).
The corresponding H$\beta$ luminosity is in agreement 
with the observed value (see Section 6.1).

\section{Modeling the UV-optical continuum}

The equivalent width (Ew) of H$\beta$ measured in the spectrum {\it b}, 7~\AA, is lower than the Ew reported from
ground-based observations, 34~--~36~\AA\ (Vacca \& Conti 1992; Schaerer et al. 1999; Guseva et al. 2000), and much lower 
than the predicted Ew(H$\beta$) for 
an instantaneous burst between 3~Myr (Ew(H$\beta$)~=~200~\AA) and 4~Myr (Ew(H$\beta$)~=~63~\AA)
 at solar metallicity, for a Salpeter IMF
with \mup~= ~100~M$_\odot$. This large discrepancy may be caused by an additional stellar population that contributes
to the optical continuum flux but not to the ionization. Certainly, the bulge population may contribute significantly
to the continuum flux detected in the ground-based observations of 
Guseva el al.  obtained through a slit of 2~arcsec width. 

 We use  
the SED of the central cluster of NGC 3049 over the range of $\sim$1200~--~6800~\AA\ to
find evidence for an additional older stellar population that might dilute the  Ew(H$\beta$).
However, with the exception of the WR features, we 
have not detected any other stellar lines at optical wavelengths. The Balmer absorption
lines, or the CaII H and K lines,
which could indicate a population of B-A stars, are not detected in the spectrum. Thus, this featureless
 blue optical continuum
emitted by the central 0.1$\times$2~arcsec could be provided mainly by a single age, young stellar population. Even so,
we have compared the shape of the continuum with the models as a further constraint on the stellar population 
providing the optical continuum flux of the central spectrum.

Although the images and spectra indicate a very compact central cluster
(FWHM of the profiles $\sim$0.12~--~0.14~arcsec),
the optical continuum flux measured through a slit of 0.1~arcsec may represent only some fraction of the total 
optical continuum flux
emitted by the cluster. To estimate this fraction, we have compared the optical continuum flux measured 
in the spectrum {\it b} 
(corresponding to the central 0.5$\times$0.1 arcsec; $2.9\times10^{-16}$~\ergsa\ at 6900~\AA)
with the flux measured in the optical image in an aperture of 0.5$\times$0.5~arcsec 
($3.4\times10^{-16}$~erg~s$^{-1}$~cm$^{-2}$~\AA$^{-1}$).
A direct comparison of these 
values indicates that the extracted continuum flux fraction through the central 0.5$\times$0.1~arcsec 
spectrum is about 85\% of the total flux. The optical continuum flux of the spectrum {\it b} has 
to be scaled by a factor of
1.15\footnote{Note that the $L_{4660}$ estimated in the
previous section has to be multiplied by this fraction as well to account 
for the total flux} 
for a comparison with the UV continuum. Taking into account this scaling factor, we have merged 
the UV and optical continua 
corresponding to the spectrum {\it b}.     
We have also corrected the continuum flux distribution for reddening with
the Calzetti et al. (2000) obscuration law and $E(B-V)=0.2$.
We find that the dereddened continuum is very well fitted by the SED of an instantaneous burst of
age 3~Myr and formed following a
Salpeter IMF with \mup~=~100~M$_\odot$. No additional older stellar population is required to fit
 the continuum, and the
stellar population must contribute very little to the flux detected through a 0.1~arcsec slit. 
This result contrasts with those
obtained based on the analysis of ground-based observations (obtained with a slit width of several arcsec),
 that claim that a post-starburst stellar population contributes significantly
to the optical continuum (Mas-Hesse \& Kunth 1999; Schaerer et al. 2000). Probably, this 
post-starburst stellar population is more spatially extended than the ionizing stars, therefore, 
it does not contribute significantly to the continuum flux detected in the STIS narrow slit.

The contribution from an additional stellar population to the young burst is not the explanation 
of the large discrepancy between the STIS observed EW(H$\beta$) and the value predicted by the 3~--~4~Myr instantaneous
burst models. However, the difference can be explained if the ionized gas is more extended than the continuum flux. 
If the total H$\beta$ flux measured from the ground-based observations (scale of $\sim$100~pc)
is divided by the optical continuum flux
near H$\beta$ measured in the spectrum {\it b} (scale of a few pc),
 the Ew(H$\beta$) is 120~\AA, which is in perfect agreement with the predicted values
for a 3.5~Myr instantaneous burst. There is no need for an extended burst or an older stellar population in the central
0.5$\times$0.1~arcsec.

\section{Modeling the nebular lines}

It is well known that starbursts display a spectral dichotomy. As we have described, the UV wavelengths
are dominated by wind and photospheric absorption lines from massive stars, and the optical wavelengths by
nebular emission lines. Because the stellar winds of non-WR stars
are optically thin to the ionizing photons from massive stars, the photons
escape and are absorbed by the surrounding interstellar medium. Then, the ionized gas cools down via radiation from the emission
lines. Thus, like the stellar UV lines, the emission lines depend on the stellar content of the starburst.
However, it is more difficult to constrain the age, IMF, and star-formation history of the starburst 
through the nebular lines because 
the modeling of the emission lines is a degenerate problem. In addition to the radiation field, the
strength of these lines depends on the chemical composition, the density structure, 
and the geometry of the ionized gas.
Even so, photoionization models of various starbursts and HII regions can successfully
constrain the stellar content
(e.g., Garc\'\i a-Vargas et al. 1997 for the HII regions of NGC~7714;
Gonz\'alez Delgado et al. 1999 for the nuclear starburst of NGC~7714; Luridiana et al. 1999 for NGC~2363;
Stasi\'{n}ska \& Schaerer 1999 for IZw18;
Gonz\'alez Delgado \& P\'erez 2000 for NGC~604;  
Luridiana \& Peimbert 2001 for NGC~5461).

In the present study
 we perform a critical test of the inputs to the photoionization models, in particular the radiation 
field of the ionizing stars. The stellar ionizing spectrum is known a priori
from the analysis of the UV spectra. Therefore we expect to constrain the IMF and to check if 
the stellar and the nebular analysis in this high-metallicity starburst lead to similar 
\mup. We begin with a description of the observational constraints 
and the input parameters for the photoionization models.

\subsection{Observational constraints}

The observed quantities to constrain the photoionization models are:

{\em Geometry of the HII region}. The [OIII] and H$\alpha$ images indicate that the region is very compact
with a radius of $\sim$1.2~arcsec, corresponding to $\sim$100~pc. A sphere with a Stromgren radius of $\sim$100~pc
is a simple geometry that seems adequate to describe the ionized region.

{\em Hydrogen recombination lines}. The Balmer line intensities depend on the ionizing photon luminosity, $Q$(H).
We have estimated $Q$(H) from the mass of the cluster, which is derived from the luminosity (see Table~2). 
This $Q$(H)
gives an H$\beta$ luminosity of log~$L$(H$\beta$)~$\approx$~40.1 (erg~s$^{-1}$). 
This luminosity agrees well with the measurements
of Vacca \& Conti (1992) (39.9~erg~s$^{-1}$) and Schaerer et al. (1999) (39.8~erg~s$^{-1}$) if 
we take into account that the narrow long-slit used for the spectroscopic observations,
 which represents a lower limit to 
the total H$\beta$ flux of the ionized region. On the other hand, Guseva et al. (2000) estimate the total H$\beta$ luminosity correcting for
slit effects and seeing. They give log~$L$(H$\beta$)~$\approx$~40.3 (erg~s$^{-1}$).

{\em HeI recombination lines}. The ratio of HeI $\lambda$5876/H$\beta$
is very sensitive to the effective temperature of the stars if $T_{\rm{eff}} \leq 40000$~K. 
Because this ratio depends
only on the ratio of the He to H ionizing photons, it must be predicted very accurately. Given the importance
of fitting this ratio
correctly, we impose that the theoretical and observed ratios must agree within less than 25\%. 
Because our spectral range does not cover the HeI $\lambda$5876 line, we took the HeI $\lambda$5876/H$\beta$ ratio from 
ground-based observations (see Table~3). After inspecting the literature, 
we consider the ratio of Guseva et al. as the 
most credible value because they have taken into account the absorption by the stellar population.

{\em Collisionally excited lines}. Ratios of forbidden over Balmer lines, such as 
[OIII]$\lambda$5007/H$\beta$, [OII]$\lambda$3727/H$\beta$,
[OI]$\lambda$6300/H$\beta$, [NII]$\lambda$6584/H$\beta$, and [SII]$\lambda$6717+6731/H$\beta$
depend on the ionization structure on the gas and on the electron temperature. The most useful 
ratios are those sensitive to the effective temperature of the stars, 
such as [OIII]$\lambda$5007/[OII]$\lambda$3727.
This ratio must be predicted to within 50\%.

{\em Electron density}. $N_{\rm{e}}$ is determined by the 
[SII]$\lambda$6717/[SII]$\lambda$6731 ratio. Ground-based observations of this ratio
indicate that $N_{\rm{e}}$ is $\sim$300~cm$^{-3}$. From our STIS spectra we 
measure $\sim$500~cm$^{-3}$ in the central 0.3~arcsec.

Table~3 gives all the observed emission~line ratios and  other additional constraints for
 the models.

\subsection{Input parameters and assumptions}

We have computed several sets of models using different 
  SEDs as the ionizing radiation field in two 
different photoionization codes, PHOTO (in the version described in Stasi\'{n}ska \& Leitherer 1996) 
and Cloudy (Ferland 1997, version C96.b2). The reason for using two codes is to perform   
a consistency check of the results. Even though 
photoionization codes have been shown before to give similar results at subsolar metallicity 
(references in Ferland \& Savin 2001), the results may diverge at  
supersolar metallicity due to numerical and physical instabilities  
under very extreme conditions.  
However, we find the same general conclusion from both photoionization codes, 
i.e., the inability to fit the  
observed emission line ratios with clusters formed 3~--~4~Myr ago containing 
 stars more massive than \mup~$\geq60$~M$_\odot$ 
(see below). 
 
The shape of the stellar radiation field (see Section~6.5 for details),
the ionizing photon luminosity, the electron density, the inner radius of the sphere, 
the filling factor, and the chemical composition are the input parameters for the 
photoionization models to predict the intensity of the emission lines.
We always assume that the gas is ionization bounded, and we adopt a fixed
ionizing photon luminosity,  $\log Q= 52.45$~ (ph~s$^{-1}$),
 as estimated from the UV continuum luminosity of the central starburst. 
The models reproduce the Balmer recombination lines with this value of $Q$. 
The electron
density is constant and equal to 300~cm$^{-3}$.

The central cluster in our STIS images of NGC~3049 is unresolved. 
We assume a typical cluster size of a few pc, the value derived for young
super star clusters in starburst galaxies (Whitmore 2002).
The FWHM of the spatial profiles in the UV and optical
continuum is $\sim$9~pc; therefore the size should be smaller than 9~pc. 
We adopt 3~pc for the inner radius of the gaseous sphere,
but the results would be similar for any value of the inner radius smaller or similar than that. 

We assume that the gas is uniformly distributed in small clumps of constant density
over the nebular volume and occupies a fraction   $\phi$ of the total volume.
A change of $\phi$
 is equivalent to a change of the ionization parameter $U$, defined as  
$Q/(4\pi R N_{\rm e} c$), where $Q$ is the ionizing photon luminosity, 
$N_{\rm e}$ the electron density, $c$ the speed of light and $R$ the outer radius of the nebula.
For a spherical geometry, the average $U$ 
is proportional to $(\phi^2N_{\rm e}Q)^{1/3}$. In our models, $\phi$ is a free parameter ranging from $\log \phi=-1.5$ to --4.0 with a step of 0.5. However, this parameter is later fixed at
 the value that provides a radius of the ionizing front equal to
the size of the HII region, $\sim$100~pc. Given the relationship between $\phi(U)$
and the ratio [SII] $\lambda$6717+6731/H$\beta$
(Gonz\'alez Delgado et al. 1999; Gonz\'alez Delgado \& P\'erez 2000), the value 
of $\phi(U)$ has to fit the [SII]/H$\beta$ ratio as well.
A filling factor ranging from 10$^{-2}$ to 10$^{-2.5}$ fits the observational constraints, 
and predicts a radius that is always within 100$\pm$20$\%$ pc.

The radiation field used as input to PHOTO is the SED  
of the evolutionary synthesis models of Schaerer 
\& Vacca (1998; SV98). They are based on the non-rotating Geneva stellar 
evolution models with the high mass-loss tracks of Meynet et al. (1994). The 
SED is built with the CoStar (Schaerer \& de Koter 1997) stellar atmosphere 
models for massive main-sequence stars, which take into account the effects of 
stellar winds, non-LTE, and line-blanketing. For WR stars, we use 
the pure He models of Schmutz, Leitherer, \&  Gruenwald (1992), and the 
plane-parallel LTE models of Kurucz (1993) for the remaining stars that 
contribute to the continuum. 
 
Different radiation fields are used as input to Cloudy. They are the SEDs from 
the SV98 models, and from SB99 (Leitherer et al. 1999). SB99 includes the same 
stellar evolutionary tracks that SV98, but different stellar atmosphere 
models. The SB99 code allows users to generate the SED with: 
i) the stellar atmospheres grid compiled by Lejeune et al. (1997), 
supplemented with the Schmutz et al. (1992) WR-star models; or 
ii) the Kurucz (1993) stellar atmosphere models for all stars, 
including WR stars. 
 SEDs with option i) constitute our standard models, but we 
ran additional models with option ii) to investigate how our conclusions 
depend on the stellar atmospheres.

Even though our UV analysis favors the instantaneous burst scenario, the SEDs 
used here assume two different star-formation histories: 
instantaneous and continuous star formation at a constant rate. We assume that 
the slope of the IMF is Salpeter ($\alpha= 2.35$), and the upper mass cut-off 
is set to 40, 60 and 100~M$_\odot$. We also assume that the stars evolve from 
the main sequence following the solar ($Z=0.02$) and twice solar ($Z=0.04$)  
metallicity stellar 
tracks.

The chemical composition of the gas is also a free parameter in the models. 
There is strong evidence in favor of supersolar
metallicity for stars and ionized gas; however, considering the difficulty of making a good  abundance estimate in ionized regions
where [OIII] $\lambda$4363 is not detected, we take the oxygen abundance as a free parameter that changes from half solar
to twice solar (0.5, 0.75, 1, 1.5 and 2~Z$_\odot$) to explore a large range of possible solutions. We scale the other elements following 
the prescription of McGaugh (1991). However, we generated
an extended  set models to analyze the effect of dust, in which the abundance of 
the elements are scaled following the ratios given in Table~14 of the Hazy I Cloudy manual. The effect of dust could be
important in high-metallicity HII regions
(Shields \& Kennicutt 1995). To further check that our conclusions are not
dependent on dust effects, we have
 computed additional Cloudy models with:\\
i) Depletion of the gas-phase composition. A fraction of the abundance of the elements 
may be in a grain phase; thus, the gas phase abundance of the elements is depleted. 
No absorption or heating by dust are allowed. The depletion factors
are those listed in Table~16 of the Hazy~I Cloudy manual (Ferland 1997).
These factors are based
on the depletions listed by Jenkins (1987) and Cowie \& Songaila (1986).\\
ii) Depletion of the gas-phase composition plus absorption by grains. Models 
include grains composed of graphite, silicate, and polycyclic aromatic 
hydrocarbon (PAH). The 
properties of the grains are listed in Table~17 of the Hazy~I
 Cloudy manual (Ferland 1997). Heating and cooling by dust grains are not allowed.\\
iii) Depletion of the gas-phase composition, plus absorption, and heating and cooling by dust grains.

First (Section~6.3), we discuss the results from the standard models (Cloudy+SB99 with stellar atmosphere option i), then (Section~6.4) 
the effect of dust on the emission-line ratios, and finally (Section~6.5)
the models using different stellar atmospheres.

\subsection{Results from the standard models}

Figure~12 shows the HeI $\lambda$5876/H$\beta$, [OIII]/[OII], [OIII]/H$\beta$ and [SII]/H$\beta$ as a 
function of the age for instantaneous burst models with a Salpeter IMF and 
\mup~=~100~M$_\odot$. The different symbols
correspond to different metallicities of the gas, 
but we always assume that the stars evolve from the main sequence 
following the solar tracks ($Z=0.02$). These models do not
 fit the observations if the cluster is 3~--~4~Myr old, and if the IMF is Salpeter-like 
with \mup~=~100~M$_\odot$ because 
the high-excitation emission-line and HeI/H$\beta$ ratios are higher than observed. 
The observed ratios suggest
that the ionizing radiation field has to be softer than the extreme ultraviolet field predicted by the solar metallicity models.

The emission-line properties of the nebula depend strongly on the electron temperature of the ionized gas. We can modify
the thermal balance of the region by changing the gas metallicity. However, changing the metallicity does not help. 
The HeI/H$\beta$ and [OIII]/H$\beta$ line ratios point out to contradicting solutions.
HeI/H$\beta$ requires a low metallicity ($\leq 0.5$~Z$_\odot$), while [OIII]/H$\beta$ and [OIII]/[OII] indicate a high metallicity ($\geq  2$~Z$_\odot$) of the ionized gas. Consequently, ionized  gas with a metallicity 
lower than the solar value is not a possible
solution.
Alternatively, the observed emission-line ratios can be accounted for by
the solar metallicity models if the cluster is $\sim$2~Myr old. However, the UV  wind lines completely
rule out this solution (Figure~10).

Therefore, instantaneous bursts with a Salpeter IMF slope and \mup~=~100~M$_\odot$ 
cannot fit the observed ratios because
the radiation field predicted is harder than the radiation seen by the gas. 
The softness of the SED can be understood if
a lower fraction of massive stars is formed in the cluster. Because the photoionization between 3~--~5 Myr is strongly affected by
WR stars, the radiation field becomes significantly softer
 if the number of WR stars
decreases considerably with respect to the O stars. The WR/O ratio 
can be changed by decreasing \mup. Figure~13 plots
the resulting emission-line ratios as a function of the age for 
\mup~=~100, 60 and 40~M$_\odot$. 

Only models with \mup~=~40~M$_\odot$ (~60~M$_\odot$) and age $\sim$2.5~Myr ($\sim$2-2.5~Myr) 
are close to the observed values. However, a cluster formed with this \mup\ (\mup~=~40~M$_\odot$)
 will also
has a very low fraction of O and blue supergiant stars. Clusters younger than 2.5~--~3~Myr 
can be ruled out because blue supergiant stars are clearly contributing to the UV continuum.
Therefore, the predicted wind lines are very weak in comparison with
the observed lines (Figure~10). Moreover, models with a truncated IMF would
underpredict the number of WR stars with respect to the observations.

We have also computed models with SEDs for continuous star formation. 
Again, models with a large fraction of 
massive stars do not fit the observed emission-line ratios (Figure~14). 
A solution can be  obtained if the extreme ultraviolet
is built with a low fraction of WR stars.  Models with \mup~=~40 M$_\odot$ and
constant star formation predict correct HeI/H$\beta$ 
and collisional emission-line ratios. However, this solution is not consistent with the UV wind lines
and the WR numbers.

Similar conclusions are obtained if the SED is built with the $Z=0.04$
 (2~Z$_\odot$) stellar evolutionary tracks. 
Again, we cannot find a consistent solution to fit the nebular emission lines and UV wind lines. The extreme UV
SED predicted by the evolutionary models containing a 
large fraction of massive stars, and in particular WR stars,
is much harder than the radiation ionizing field seen by the gas.
As a result, the photoionization models are not able to
predict the low excitation observed in the high-metallicity HII region of NGC 3049.

Although these conclusions are based on the hypothesis that the region is
ionization bounded, the result 
about the inconsistency between UV wind lines and nebular lines does not depend on this hypothesis. 
Assuming a density bounded HII region does not resolve this inconsistency. 
It is not certain a priori that there is sufficient nebular matter 
surrounding the stellar cluster to absorb all the ionizing photons in HII regions. 
Evidence that some HII regions may be density bounded has been 
reported previously (e.g., Beckman et al. 2000; Rela\~no \& Peimbert 2002; 
Castellanos, D\'\i az \& Tenorio-Tagle 2002).  
However, we note that the H$\beta$ luminosity obtained under the assumption 
that all the Lyman photons are absorbed by the nebular gas is 
compatible with the observed value in NGC~3049. Leaving some room for 
uncertainties, one can predict the verdict from density 
bounded models. In order to reproduce the observed size of the HII 
region, density bounded models should have a slightly different 
density and filling factor than taken for the ionization bounded 
models. Whatever assumption is made for the density bounded models 
(for example, one could construct them  to reproduce the observed 
[OIII]/[OII] ratio), their HeI/H$\beta$ ratio will be equal to or larger than 
predicted by ionization bounded models. Therefore, our conclusions 
about the inconsistency between UV wind lines and nebular lines are
robust in this respect.

\subsection{Results from the dusty models}

High-metallicity HII regions are regions
 where a significant amount of dust can exist. Because the electron temperature is low,
dust easily produces important changes to the thermal structure of the HII region, and 
therefore to the emergent optical 
emission-line spectrum. In this section, we investigate
 if dusty models are able to fit the observed emission-line ratios in NGC~3049.

Dust can modify the HII thermal structure in several ways: 1) The metals
in the gas phase can be depleted onto grains. This depletion tends to increase the electron temperature of the ionized region. 2) Grains
can selectively absorb a fraction of the ionizing photons; in this way
they modify the radiation field seen by the gas.
3) Grains can provide photoelectric heating of the gas; subsequently the gas
 can cool down by electron capture. 
The consequences of dust in metal-rich HII regions have been discussed
 by Shields \& Kennicutt (1995).
They found that depletion of the gas-phase abundances introduces the strongest perturbation on the optical spectrum,
and grain heating can be particularly important for enhancing the high-ionization lines.

Because dust can also modify $Q$(He)/$Q$(H), we have computed models using 
the SED from SB99 for instantaneous
burst models with a Salpeter IMF and \mup~=~100~M$_\odot$ and assuming: 
i) only depletion of the gas-phase
abundance and ii) absorption and heating by grains. For an age of
 3~--~4~Myr, the predicted HeI $\lambda$5876/H$\beta$
is $\leq$10\% lower than the ratio predicted by dust-free models, but the collisional emission ratios are higher.
Thus, dust cannot reconcile the observed and the predicted emission-line ratios 
(Figure~15).

\subsection{Results using SEDs with different stellar atmospheres}

Our previous analysis suggests the failure of the SEDs to predict the
extreme UV radiation field which is required to reproduce the observed emission 
lines. SB99 is optimized for 
young star forming regions and has been successfully used for consistent models at subsolar metallicity. Therefore it seems likely
that the predicted SED in the WR  phase is too hard in high-metallicity clusters. We have computed 
additional solar metallicity dust-free instantaneous 
burst models with i) Kurucz atmospheres in {\em all} phases  and ii) 
CoStar atmospheres for O and Schmutz atmospheres for WR stars. This allows us 
to  check if the problem 
is related to the pure-He WR stellar atmosphere models.

At solar metallicity, models using CoStar predicts roughly the same HeI $\lambda$5876/H$\beta$ as SB99. 
The high-excitation lines are also similar
to SB99 for the Wolf-Rayet phase, but CoStar predicts larger ratios at younger age due to the inclusion of the wind
blanketing in the atmospheres of the main-sequence stars. 
The agreement at 3~--~5~Myr is not surprising because CoStar and 
SB99 both use the Schmutz unblanketed
WR atmosphere. Thus, CoStar models also fail to reproduce the observed emission-line ratios
 in NGC~3049 (Figure~16). 
However, SB99 with Kurucz models predict emission~line ratios that are
 similar to the prediction of the standard SB99 models
with \mup~=~40~M$_\odot$ (Figure~16). Although these models fit
the collisional lines, they underpredict the  HeI $\lambda$5876/H$\beta$
if the cluster is 3~--~4~Myr old. This suggests that the static
Kurucz atmospheres are equally inadequate for modeling the WR fluxes.
This is well known since the pioneering work of Hillier (1987).

We conclude that pure-He atmosphere models are not adequate for predicting
HeI $\lambda$5876/H$\beta$ and the high-excitation emission-line ratios in
high-metallicity HII regions. The new generation of WR atmospheres with
line-blanketing calculated by Smith, Norris, \& Crowther (2002) can potentially 
solve this problem. These atmospheres include the most important
metals and have a different density and temperature structure in comparison
with the Schmutz models. When coupled with Starburst99, the models of Smith
et al. give a softer far-UV radiation field at high metallicity. This is the
result of the combination of three effects: i) inclusion of blanketing
suppresses the far-UV flux; ii) a higher wind density leads to recombination
shortward of 228~\AA; iii) the WR core temperature of the evolutionary
tracks was found to be too high to be compatible with the new atmospheres.
Smith et al find that effect i) is more important for late WR stars, whereas
effect ii) dominates for early WR stars. The third effect is the result of
different temperature definitions in the evolution and atmosphere models. It
can only be calibrated empirically, e.g., by comparing the emergent
radiation field to observations. Smith et al. performed an initial
calibration and provide a recipe for the adjustment of the evolutionary
temperatures. 
The new models must produce less ionizing radiation at
frequencies above the HeII ionizing limit than the Schmutz atmospheres in
order not to overpredict the
high-excitation ratios. On the other hand, their $Q(\rm{He})$/$Q(\rm{H})$ must
be high enough in order not to underestimate HeI $\lambda$5876/H$\beta$ as
the Kurucz atmospheres do.
A new version of SB99 including the new set of the stellar atmospheres prepared
by Smith et al. was released on 2002 July 23. SEDs obtained with these 
blanketed WR and O stellar atmospheres at solar and two solar metallicities
have been applied to NGC~3049. The results are shown in Figure 17. As it was expected,
the models predict lower HeI $\lambda$5876/H$\beta$ and lower high-excitation collisional lines 
than the Schmutz et al. unblanketed WR atmospheres and are closer to the observed
values. This comparison of the models with NGC~3049 largely supports the view that the softening 
of the far-UV radiation field at high metallicity is not only a consequence of the 
inclusion the line-blanketing effect but also of the coupling between 
the WR temperature of the evolutionary tracks and the new atmospheres.
Future tests on high metallicity HII regions will have to confirm and refine the calibration 
between these two different temperatures proposed by Smith et al.

\section{Summary and Conclusions}

We have obtained HST ultraviolet and optical STIS spectroscopic and
imaging observations of the metal-rich starburst NGC~3049. 
These data are interpreted using evolutionary synthesis models optimized for star forming regions, which allow us to 
constrain the stellar content of the nuclear starburst. From the analysis of the UV-optical
continuum and the optical emission lines we have obtained the following results:

\begin{itemize}

\item{The nuclear starburst in NGC~3049 is very compact.
 The UV-optical continuum is dominated by a central
cluster that is unresolved in the UV and optical STIS images. The FWHM of the spatial profiles is only $\sim$0.1~arcsec, 
corresponding to 9~pc for a distance of 18 Mpc.}

\item{The central UV flux shows strong wind (CIV, SiIV, NV) and
photospheric lines (OV $\lambda$1371, FeV $\lambda$1360-1380, SiIII $\lambda$1417, 
CIII $\lambda$1427, and SV $\lambda$1501). These lines suggest a powerful starburst. The modeling of the
continuum and stellar absorption lines indicates that 
a high-metallicity ($Z=0.02-0.04$) 3~--~4~Myr old instantaneous burst
with a Salpeter IMF and \mup~=~100~M$_\odot$ fits  the central 0.5~arcsec
 spectrum observed in the UV. 
The mass estimated for the cluster is $\sim$10$^6$~M$_\odot$, if the extinction toward the cluster 
is $E(B-V)=0.2$. The reddening-corrected luminosity of this cluster accounts
for almost half of the far-IR luminosity of NGC~3049.
The massive stars in this cluster provide enough photons 
($\log Q= 52.30$ (ph~s$^{-1}$)) to explain the observed nuclear H$\alpha$ luminosity.
Models with very few massive stars (e.g., with \mup~$\leq40$~M$_\odot$) 
are not able to fit the UV stellar lines.
The analysis of the central 1.2~arcsec UV flux also indicates that there is no significant change of the extinction and a very small age spread.   }

\item{The central 0.3~arcsec in the optical shows 
WR features at 4660~\AA. These observations confirm previous results (Schaerer et al 1999;
Guseva et al 2000) that late WN and early and late WC stars are 
the dominant WR stellar population. The total luminosity of the bump is 
$9 \times 10^{38}$~erg~s$^{-1}$ if the stellar reddening is $E(B-V)=0.2$,
and the Ew of the bump is 14~\AA.
This luminosity is provided by $\sim$275 WR stars. A  WR/O ratio 
equal to 0.065 is estimated for the cluster. This ratio and the 
Ew of the bump
is in very good agreement with the predictions of the evolutionary 
models for a 3~--~4~Myr old high-metallicity starburst. Therefore
the number of WR stars is correctly predicted by the stellar
evolution models.}

\item{We have not detected any absorption lines that could indicate the presence of
red supergiants, or an intermediate or old stellar population. 
The blue optical continuum of the central 0.5$\times$0.1~arcsec 
is well fitted by a 3~Myr old instantaneous burst with
$E(B-V)=0.2$, as predicted by the UV continuum. Thus, no additional older stellar population contributes
significantly to the nuclear optical continuum.}

\item{The Ew of the H$\beta$ lines is only 7~\AA. This is lower than the 
values of 200~--~60~\AA\ predicted by a 3~--~4~Myr instantaneous
burst model. However, this discrepancy is naturally
explained if the ionized gas is more extended than the compact cluster.}

\item{Two kinematic components are resolved in the H$\alpha$ emission line. These 
two components are extended by $\sim$30~pc
 south-west of the maximum of the continuum, and by $\sim$8~pc at $\sim$50~pc north-east.
 The second component is 
blueshifted $\sim$60~\kms\ with respect to brightest main component. 
It may represent a bubble powered by the 
winds from massive stars. Additional evidence for an outflow of
the interstellar medium comes from the blueshift
of 200~\kms\ of the UV interstellar lines with respect to the radial velocity reported by NED of
 300~\kms\ 
with respect to the nebular optical radial velocity.}

\item{The modeling of the nebular emission lines failed to constrain
 the hot stellar content of the nuclear starburst of NGC~3049.
The models which predict emission-line ratios in agreement
 with the observations contain very few massive stars. This result,
which is clearly in contradiction with the UV spectrum, would be
 in line with previous suggestions for
a depletion of massive stars in the
IMF of metal-rich starbursts. However, this conclusion is an artifact
of the failure of the population synthesis models
to predict the ionizing radiation field. Most likely, the overly hard
radiation field results from the failure of the currently
available WR atmosphere grid. The new model set of Smith et al. (2002)
includes blanketing and uses different temperature and density structures.
These models applied to NGC~3049 predict a softer far-UV radiation field and  provide a
better match to the data.

The CoStar and SB99 models predict HeI $\lambda$5876/H$\beta$ and [OIII]/[OII] ratios that are higher than observed if the ionizing 
radiation is provided by a 3~--~4~Myr old cluster formed with a Salpeter 
IMF and \mup~=~100~M$_\odot$. SEDs using only
the Kurucz atmospheres can adequately predict the collisional line ratios 
but they underpredict $Q$(He)/$Q$(H). Thus extended atmospheres are 
required but the currently available pure He WR
models are inadequate to reproduce the nebular spectrum of high-metallicity starbursts 
in the WR phase.}

\end{itemize}

{\bf Acknowledgments}

We thank Gary Ferland for kindly making his code available, Enrique P\'erez and Miguel Cervi\~no for their 
comments from a thorough reading of the paper, Daniel Schaerer, the referee, for his detailed and useful
report, and Linda Smith for sending us their paper in advance of publication, and making their new version
of SB99 available. 
This work was supported by
Spanish projects AYA-2001-3939-C03-01
and by HST grants GO-7513.01-96A from the Space Telescope Science Institute, which is operated by 
the Association of Universities for Research in Astronomy, Inc., under NASA
contract NAS5-26555.





%





%









\clearpage

\begin{deluxetable}{cccccc}

\footnotesize

\tablecaption{UV (F25RSF2) and Optical (F28$\times$50LP) aperture photometry of NGC 3049}

\tablewidth{0pt}

\tablehead{

\colhead{Radius} & \colhead{UV Flux} & \colhead{UV mag} & \colhead{Radius} & \colhead{Opt. Flux} & \colhead{Opt. mag} \nl
\colhead{(arcsec)} & \colhead{(10$^{-15}$ erg s$^{-1}$ cm$^{-2}$ \AA$^{-1}$)} & \colhead{} & 
\colhead{(arcsec)} & \colhead{(10$^{-15}$ erg s$^{-1}$ cm$^{-2}$ \AA$^{-1}$)} & \colhead{} \nl
}
\startdata

   0.123   &   3.62     &   15.00  & 0.10 & 0.18 & 18.23 \nl

   0.247   &   4.92     &   14.67  & 0.25 & 0.33 & 17.60 \nl 

   0.494   &   6.71     &   14.33  & 0.50 & 0.50 & 17.15 \nl  

   0.741   &   8.41     &   14.09  &      &      &       \nl

   0.988   &   9.39     &   13.97  & 1.0  & 0.80 & 16.64 \nl  

   1.235   &   10.11    &   13.89  &      &      &       \nl  

   1.482   &   10.54    &   13.84  & 1.5  & 1.06 & 16.34 \nl 

   1.729   &   10.89    &   13.81  &      &      &       \nl  

   1.976   &   11.25    &   13.77  & 2.0  & 1.35 & 16.07 \nl  

   2.223   &   11.55    &   13.74  &      &      &       \nl  

   2.470   &   11.94    &   13.71  & 2.5  & 1.76 & 15.79 \nl  

   3.705   &   13.17    &   13.60  & 3.5  & 2.43 & 15.44 \nl 

           &            &          & 4.0  & 2.72 & 15.31 \nl

           &            &          & 4.5  & 3.01 & 15.20 \nl

   4.940   &   13.91    &   13.54  & 5.0  & 3.38 & 15.11 \nl  

   6.175   &   14.46    &   13.50  &      &      &       \nl  

   7.410   &   14.91    &   13.47  & 7.5  & 4.39 & 14.79 \nl  

   8.645   &   15.38    &   13.43  &      &      &       \nl  

   9.880   &   15.89    &   13.40  & 10.0 & 5.24 & 14.60 \nl

\enddata

\end{deluxetable}

\begin{deluxetable}{cccccc}

\footnotesize

\tablecaption{UV properties of the different extracted one-dimensional UV spectra that are marked in plot 3 and 4.}

\tablewidth{0pt}

\tablehead{

\colhead{Spectrum} & \colhead{Extraction} & \colhead{Flux$_{1500}$} & \colhead{$\log L_{1500}$} & \colhead{Mass} 

& \colhead{$\log Q$} \nl

\colhead{} & \colhead{($\pm$ arcsec)} & \colhead{(10$^{-15}$ erg s$^{-1}$ cm$^{-2}$ \AA$^{-1}$)}
 & \colhead{(\ergsa)} & \colhead{(10$^6$ M$_\odot$)} 
& \colhead{(ph s$^{-1}$)} \nl

} 

\startdata

   {\it a} &   1.275  &  12  &  39.52  & 1.5   & 52.45 \nl

   {\it b} &   0.25   &  8.2 &  39.35  & 1     & 52.30 \nl

   {\it c} &   0.125  &  1.1 &  38.49  &   &  \nl

   {\it d} &   0.125  &  1.0 &  38.42  &   &  \nl

   {\it e} &   0.125  &  0.2 &  37.74  &   &  \nl

\enddata

\end{deluxetable}

\begin{deluxetable}{lccccc}

\footnotesize

\tablecaption{Observed emission-line ratios}

\tablewidth{0pt}

\tablehead{

\colhead{line} & \colhead{Guseva et al.} & \colhead{Schaerer et al.} & \colhead{Vacca \& Conti} & \colhead{STIS ({\it a})} 

& \colhead{STIS ({\it b})} \\

} 

\startdata

{[OII]}$\lambda$3727/H$\beta$   &   1.26     &   ---    &  1.72  & 1.1   & 1.2 \\

{[OIII]}$\lambda$5007/H$\beta$  &   0.322    &   0.338  &  0.321 & 0.36  & 0.45 \\

 HeI$\lambda$5876/H$\beta$     &   0.104    &   0.133  &  0.091 & ---   & --- \\

{[OI]}$\lambda$6300/H$\beta$    &   ---      &   0.023  &  ---   & ---   & --- \\

{[NII]}$\lambda$6584/H$\beta$   &   1.074    &   ---    &  1.08  & 1.23  & 1.31  \\

{[SII]}$\lambda$6717/H$\beta$   &   0.319    &   ---    &  0.308 & 0.21  & 0.19 \\

{[SII]}$\lambda$6731/H$\beta$   &   0.279    &   ---    &  0.259 & 0.23  & 0.22  \\

&            &          &        &       &        \\

Ew(H$\beta$) (\AA)            &   35.9     &   34     &  35.1  & 12.4  &  7 \\

log $L$(H$\beta$) (erg s$^{-1}$)       &   40.3     &  39.80   &  39.90 & ---   &  --- \\

\enddata

\end{deluxetable}

\clearpage


\begin{figure} 


\figcaption{Ground-based H$\alpha$ image observed with the JKT. The H$\alpha$ 
emission is detected along the stellar bar and in a ring of 30 arcsec radius.
North is up and east to the left. This is the orientation in all the images. }

\end{figure}


\begin{figure} 


\figcaption{HST+STIS images of NGC 3049: a) UV wavelength with the FUV-MAMA detector
and the filter F25SRF2 ($\lambda_{\rm{c}}=1469$~\AA). The sampling is 0.025~arcsec/pixel.
b) Optical wavelength with the CCD detector and the filter F28$\times$50LQ 
($\lambda_{\rm{c}}=7150$~\AA).
The sampling is 0.05 arcsec/pixel.}

\end{figure}


\begin{figure} 


\figcaption{Central 4$\times$4~arcsec emission of NGC~3049 at: a) UV wavelengths. b) Optical wavelengths. 
c) [OIII] emission line plus the underlying continuum.
The nuclear emission is dominated by a central cluster.  }

\end{figure}


\begin{figure} 

\centerline{\psfig{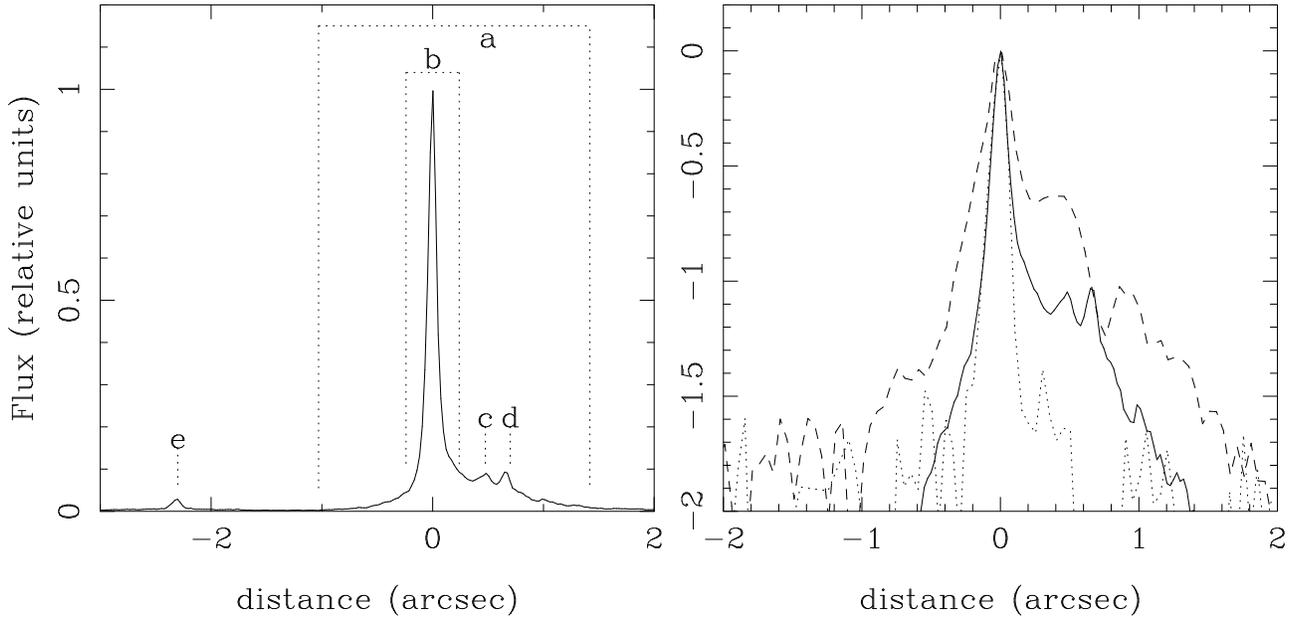}}

\figcaption{a) UV (1500 \AA) continuum flux distribution along the slit position. 
The different extracted one-dimensional spectra are marked in the plot. The flux is normalized 
to the maximum of the continuum. b) Spatial distribution of the flux (on a logarithmic scale) 
at the 1500~\AA\ UV continuum (full line), H$\alpha$ flux (dashed line) and the optical continuum
near H$\alpha$ (dotted line). These fluxes have been normalized to their maximum value at the nucleus of the galaxy. }

\end{figure}


\begin{figure}
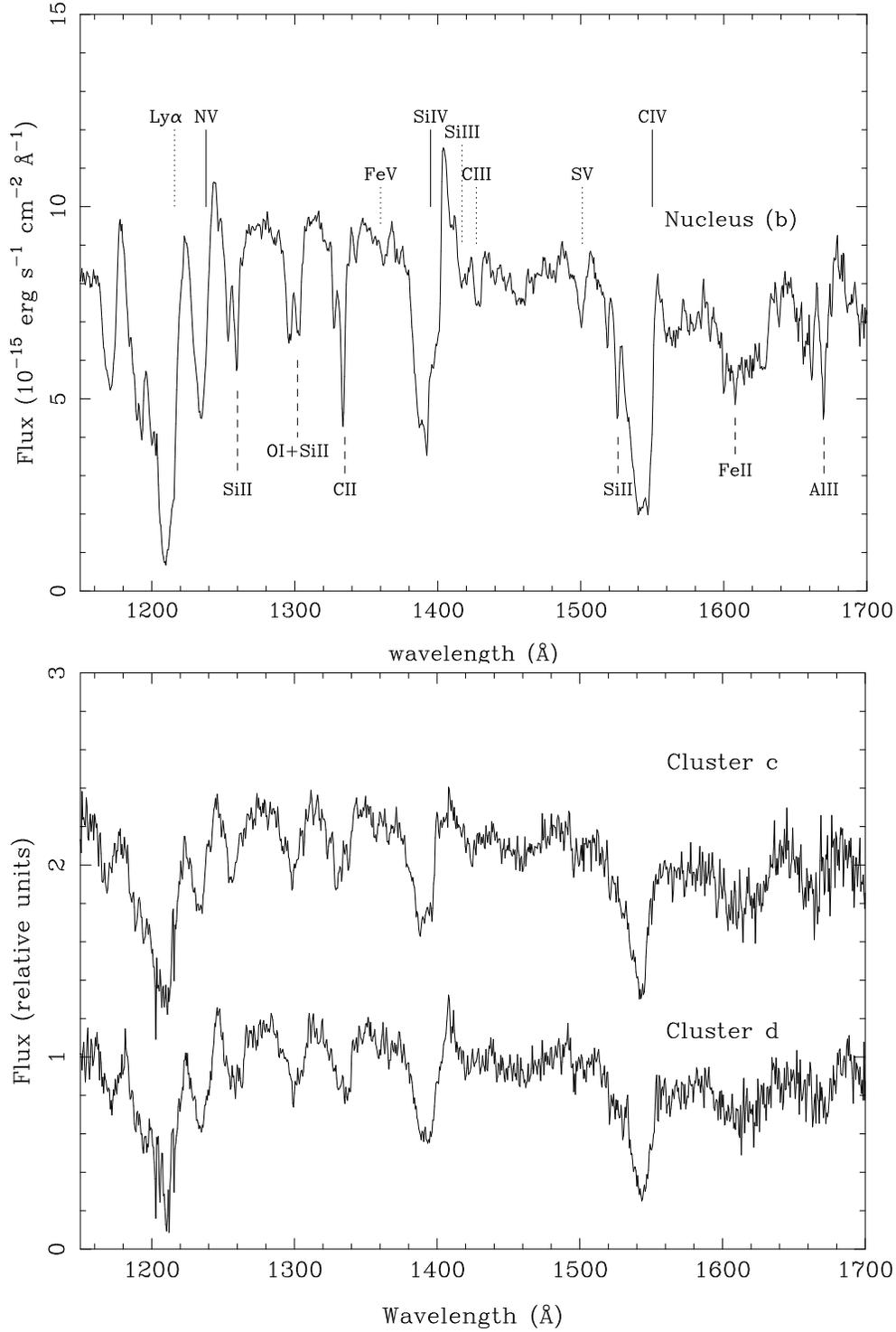
 

\centerline{\psfig{figure=f5a.ps,width=13cm,angle=270}} 

\centerline{\psfig{figure=f5b.ps,width=13cm,angle=270}} 

\figcaption{a) UV spectrum of the nucleus (marked as b in Figure~4), corresponding to the central $\pm$ 0.24~arcsec.
The most important wind (full lines), photospheric (dotted lines) and interstellar lines (dashed lines)
are marked in the plot.
b) UV spectrum of the clusters marked as c and d in the Figure~4. The flux is in 
$1\times 10^{-15}$~erg~s$^{-1}$~cm$^{-2}$~\AA$^{-1}$ units. Cluster c is shifted by 
$1\times 10^{-15}$~erg~s$^{-1}$~cm$^{-2}$~\AA$^{-1}$ in the plot.  }

\end{figure}


\begin{figure} 

\centerline{\psfig{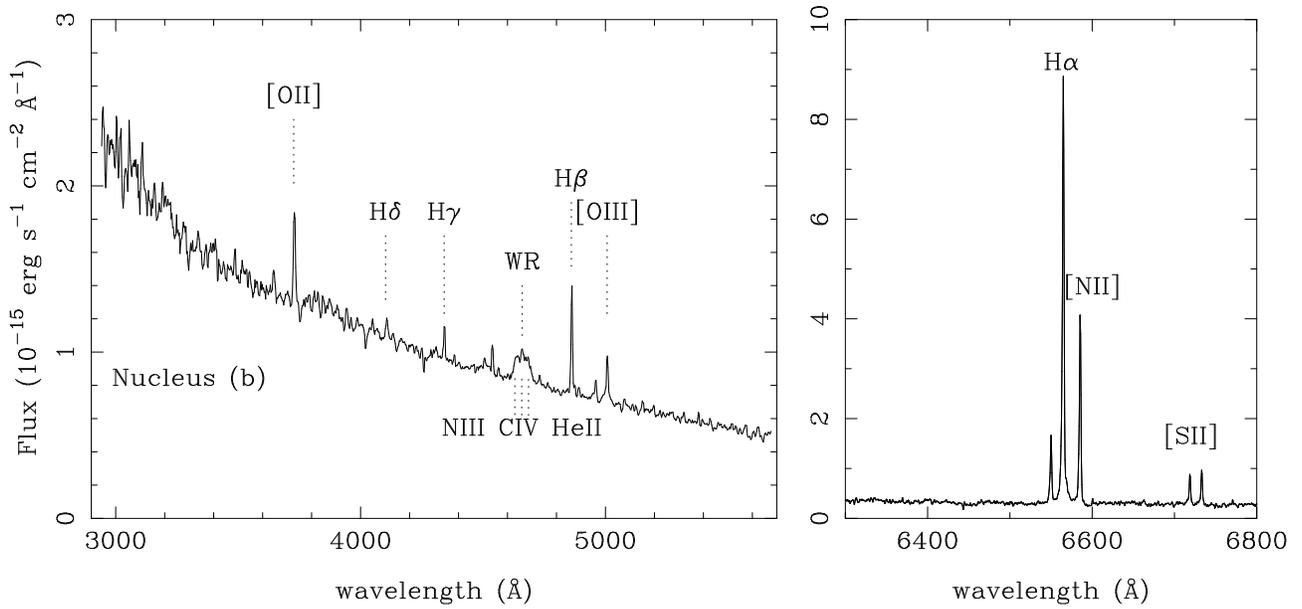}}

\figcaption{Optical continuum of the nucleus (marked as b in Figure 4) at the blue (left) and red (right) wavelength 
spectral range. The most prominent emission lines and WR features are marked in the plot. }

\end{figure}


\begin{figure} 

\centerline{\psfig{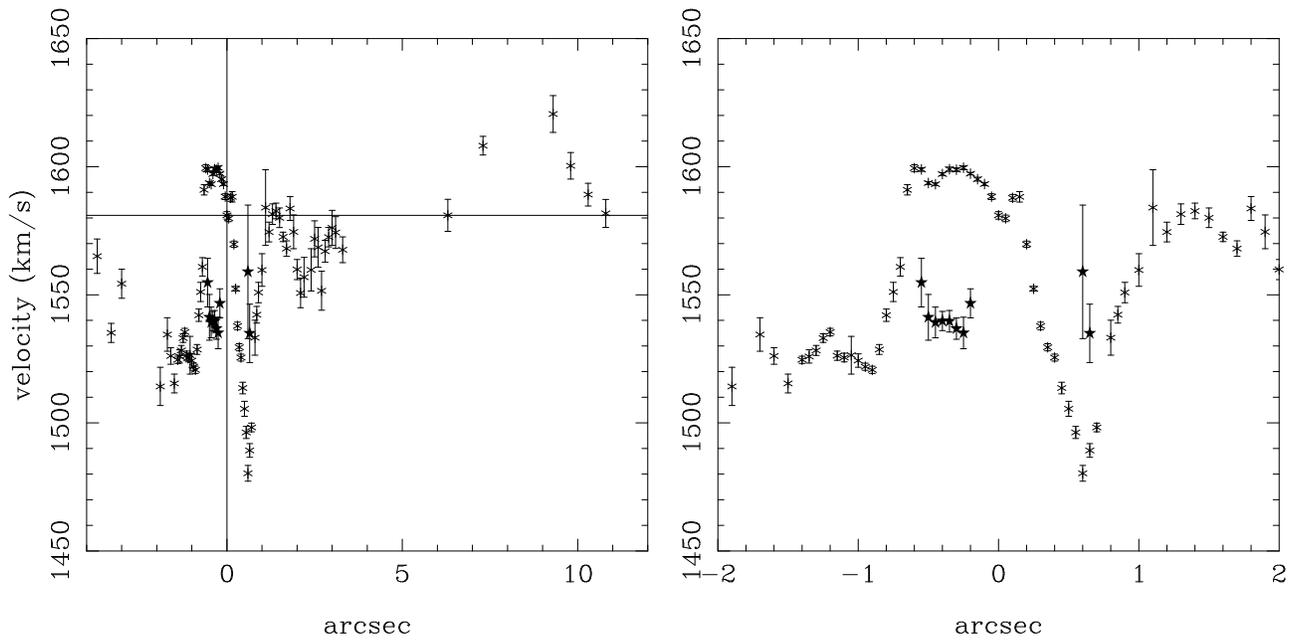}}

\figcaption{Velocity curve of the ionized gas obtained fitting Gaussians to the H$\alpha$ emission
detected along P.A.~$=-138^{\circ}$.
The second kinematic component of H$\alpha$ is plotted as stars. The velocity of the continuum peak (the nucleus)
is marked by lines. An expansion of the plot corresponding to $\pm$2~arcsec is shown on the right.  }

\end{figure}


\begin{figure} 

\centerline{\psfig{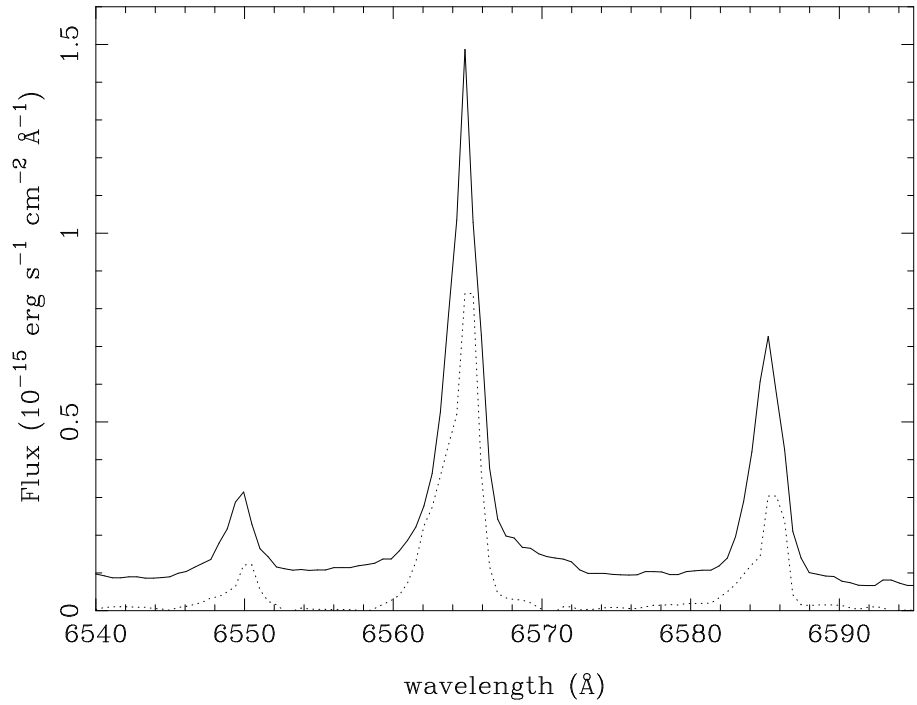}}

\figcaption{Comparison of the profile of H$\alpha$+[NII]$\lambda$6584,6548 at the pixel 
corresponding to the maximum 
of the continuum, i.e., the nucleus (full line) with the profile of these lines at 
0.35~arcsec south-west of the peak of
the continuum (dotted line). The spectrum is plotted in relative units.    }

\end{figure}


\begin{figure} 

\centerline{\psfig{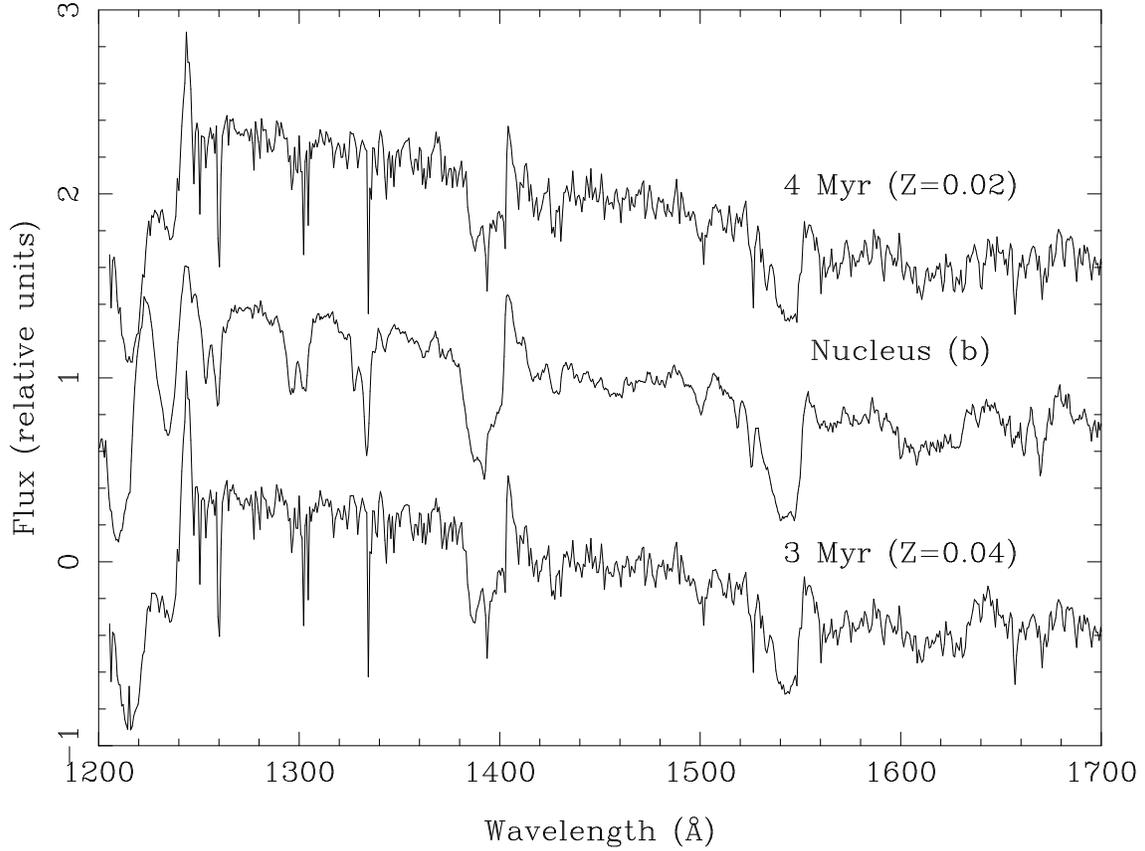}}

\figcaption{STIS/MAMA spectrum of the nucleus (b), corresponding to the UV emission of the 
central 0.5$\times$0.5 arcsec of NGC~3049, de-reddened by $E(B-V)=0.2$ 
using the Calzetti et al. (2000) 
extinction law and normalized to the corrected flux at 1500~\AA. 
The synthetic 3~Myr and 4~Myr 
(in relative units) instantaneous burst models fitting the wind lines are also plotted.
The models assume that the stars follow a Salpeter IMF with \mup~=~100 M$_\odot$ and 
evolve from the main sequence along the solar ($Z=0.02$) or the twice solar ($Z=0.04$) stellar tracks.}

\end{figure}


\begin{figure} 

\centerline{\psfig{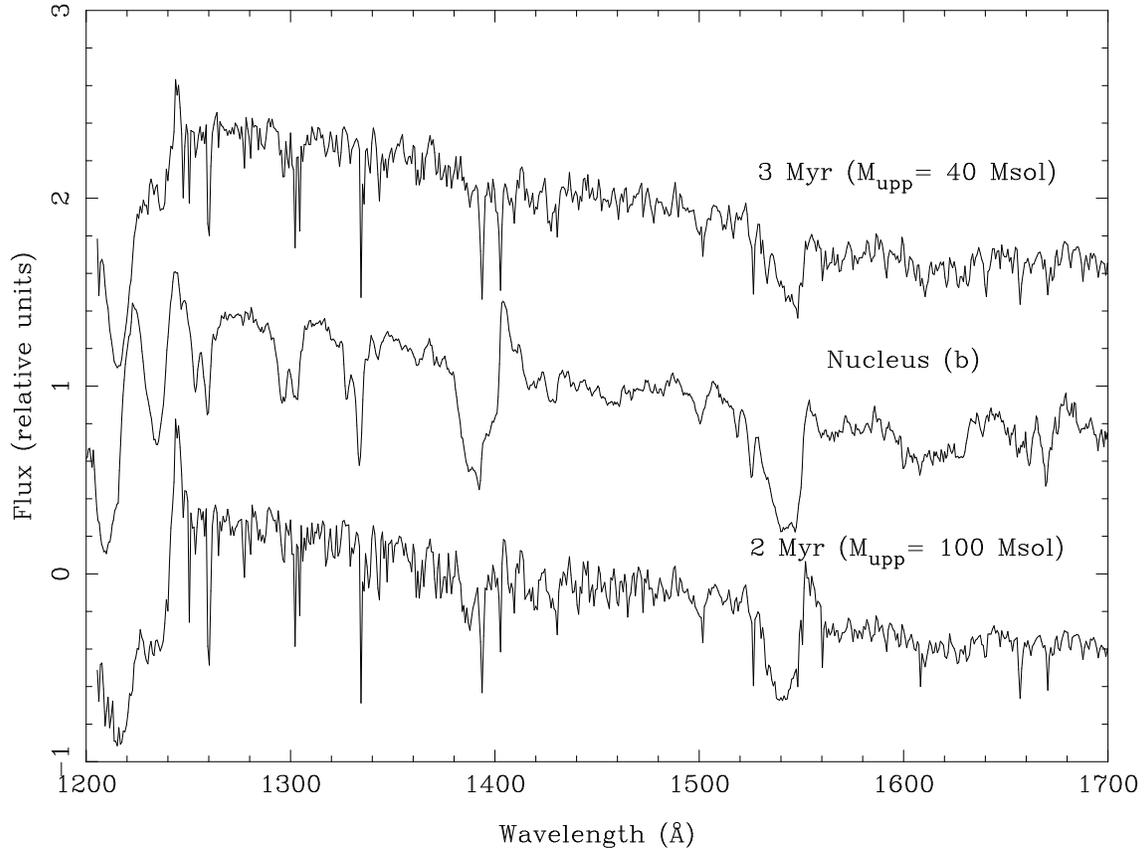}}

\figcaption{The dereddened UV nuclear spectrum is compared with synthetic 
models ($Z=0.02$). The IMF 
is Salpeter with \mup~=~100~M$_\odot$ or 40~M$_\odot$. This comparison shows that clusters
younger than $\sim$3~Myr or having a truncated upper IMF are not compatible with the observed 
UV wind lines.}

\end{figure}


\begin{figure} 

\centerline{\psfig{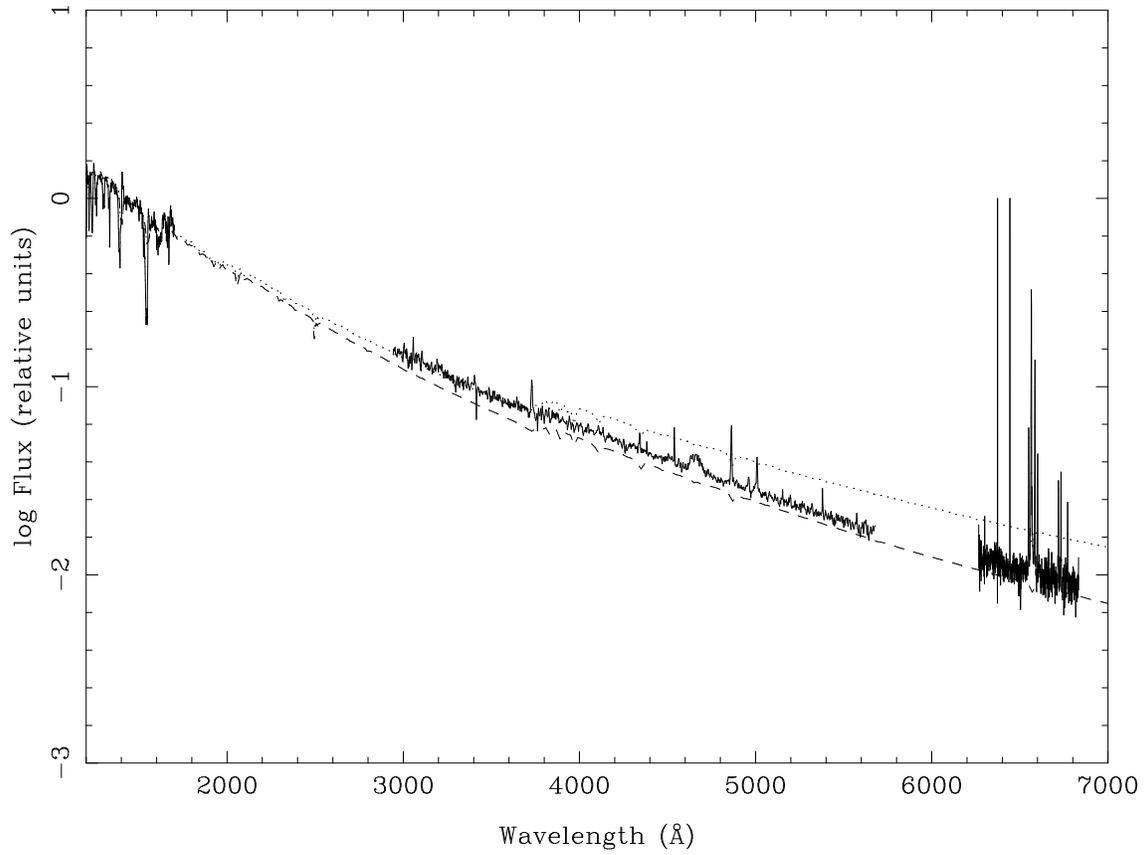}}

\figcaption{Comparison of the UV-optical continuum spectrum {\it b} normalized to the flux at 1500 
\AA\ (full line) with the SED of  a 3~Myr (dashed line) and 4~Myr (dotted line) old
 instantaneous burst model 
having Salpeter IMF, \mup~=~100~M$_\odot$, and $Z=0.02$.  }

\end{figure}


\begin{figure} 

\centerline{\psfig{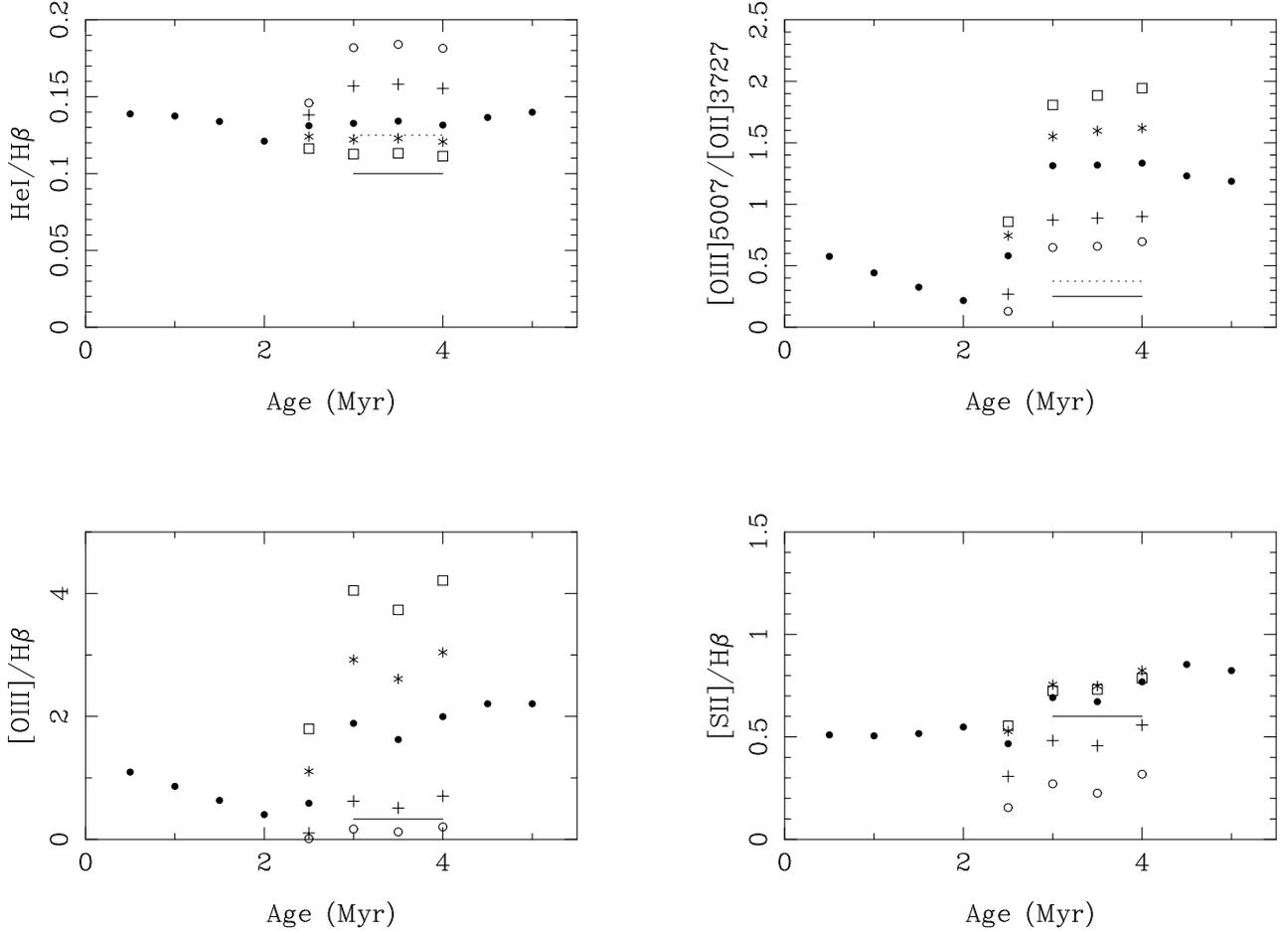}}

\figcaption{Emission-line ratios predicted as a function of age.
The observed values are indicated by a horizontal line. Dotted line is the observed value plus 
the uncertainty associated to the observed emission line ratio. The ionizing cluster formed 
in an instantaneous burst 
following a Salpeter IMF with \mup~=~100 M$_\odot$. We assume that the stars have solar metallicity. 
The oxygen abundance of the gas is 0.5~Z$_\odot$ (squares), 0.75~Z$_\odot$ (stars), Z$_\odot$ (points), 1.5~Z$_\odot$ 
(crosses) and 2~Z$_\odot$ (circles). The abundances of the other elements scale with 
McGaugh's (1991) prescription.
 The filling factor is $\log \Phi= -2.5$.}

\end{figure}


\begin{figure} 

\centerline{\psfig{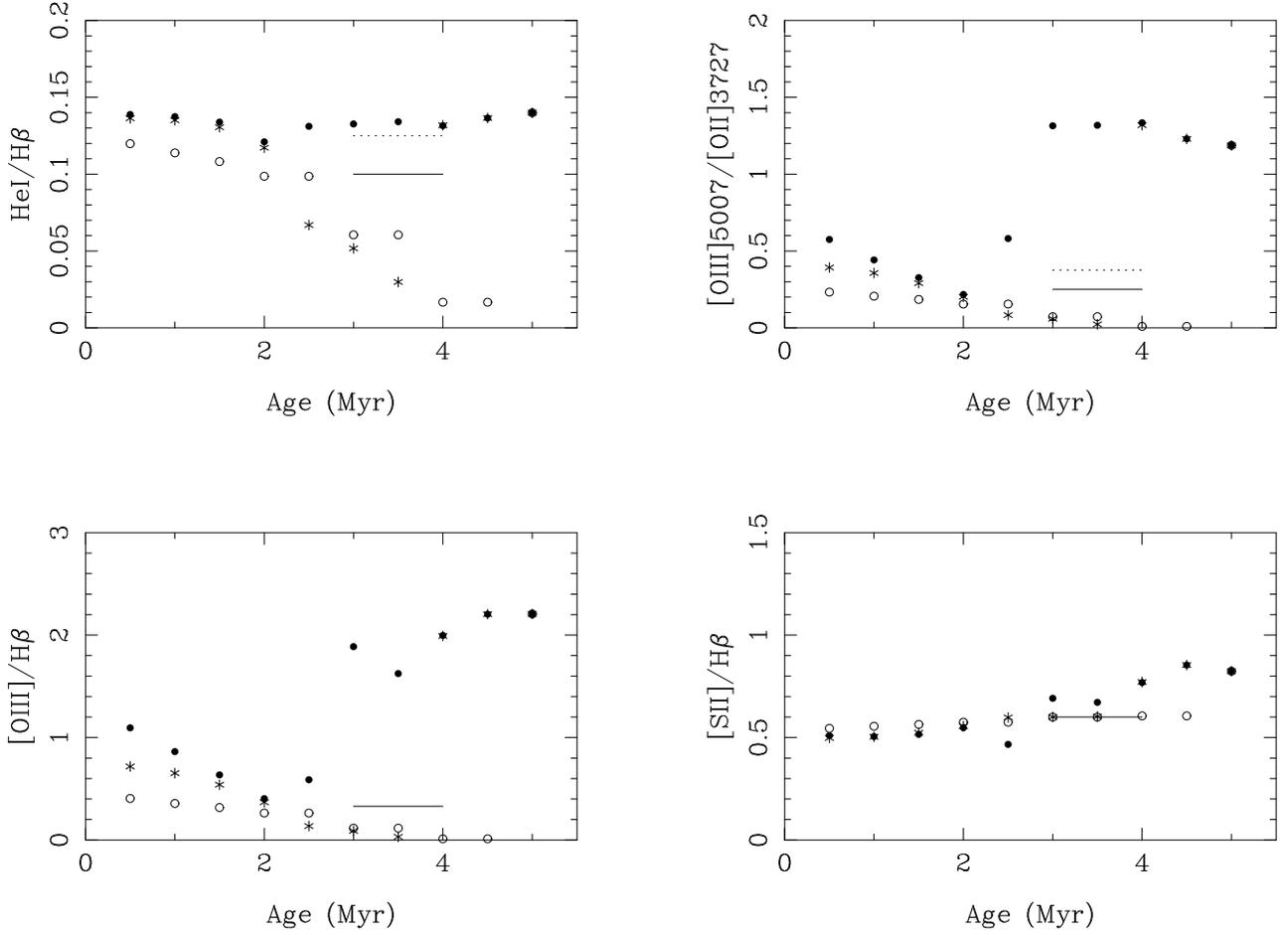}}

\figcaption{Same as Figure~12, but for clusters formed in an instantaneous burst following a Salpeter IMF with
 \mup~=~100~M$_\odot$ (points), 60~M$_\odot$ (stars), or 40~M$_\odot$ (circles).
The oxygen abundance of the gas is Z$_\odot$.  }

\end{figure}


\begin{figure} 

\centerline{\psfig{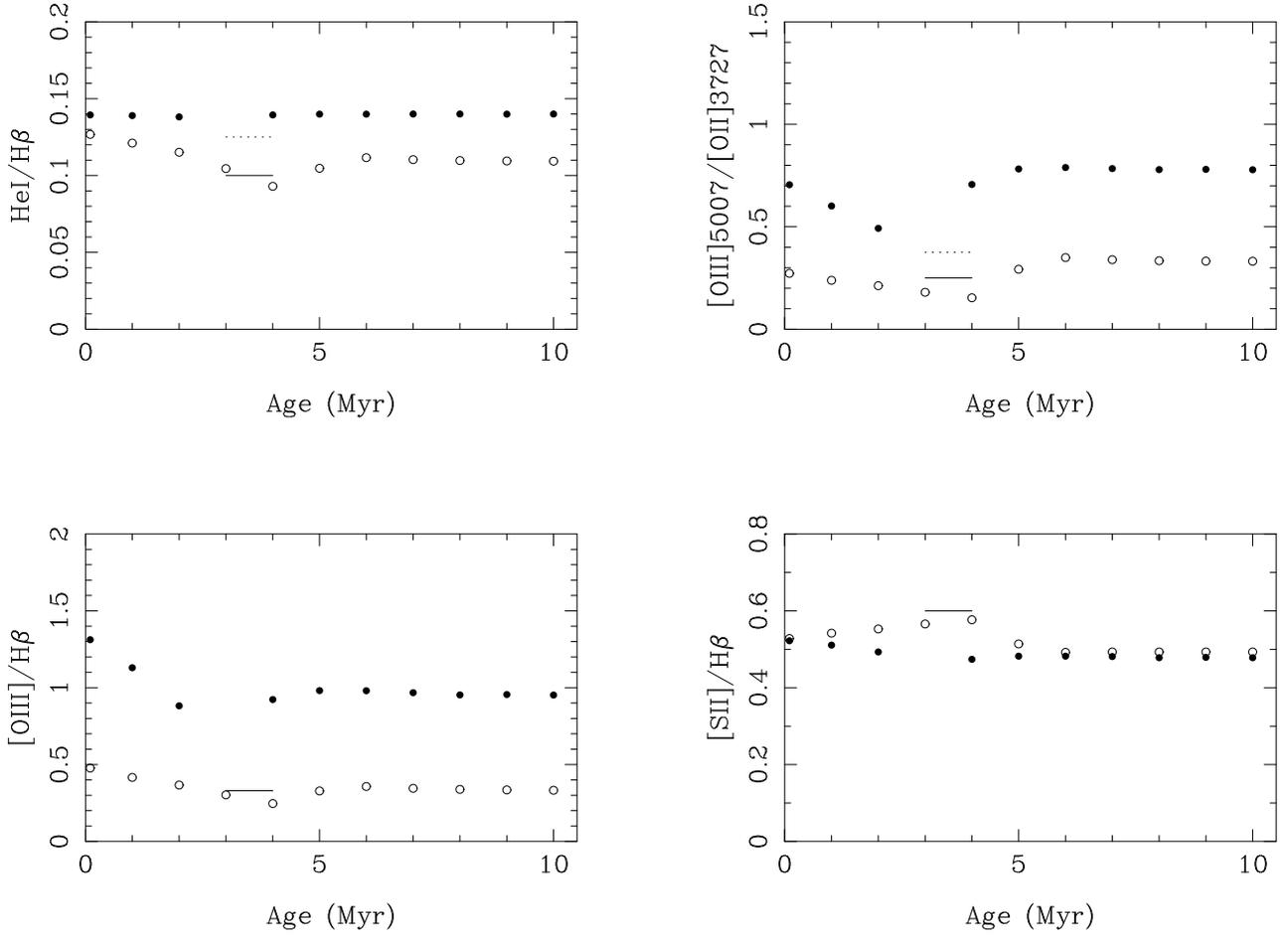}}

\figcaption{Predicted emission-line ratios for continuous star formation and Salpeter IMF with 
\mup~=~100~M$_\odot$ (points) or 40~M$_\odot$ (circles). The metallicity is solar. }

\end{figure}


\begin{figure} 

\centerline{\psfig{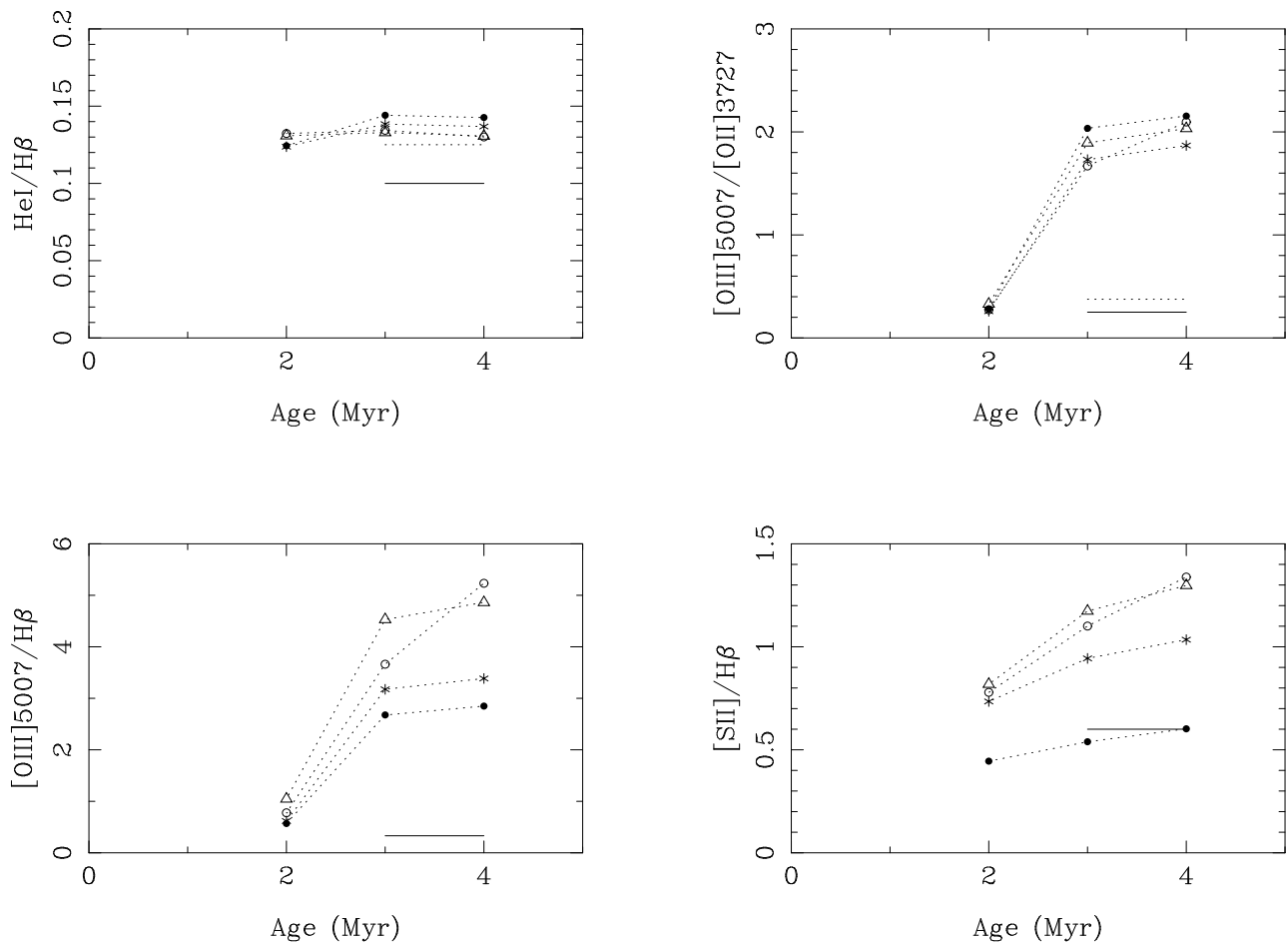}}

\figcaption{Emission-line ratios predicted assuming an instantaneous burst, Salpeter IMF, and 
\mup~=~100 M$_\odot$. The oxygen abundance of the gas is solar and the other elements scale with the solar
ratios. Several dust effects are taken into account in the models. Dust-free models are plotted by points. 
Models with depletion of the gas-phase abundances are plotted with stars, depletion and grain opacity
with circles, and depletion with grain opacity and heating + cooling with triangles.   }

\end{figure}


\begin{figure} 

\centerline{\psfig{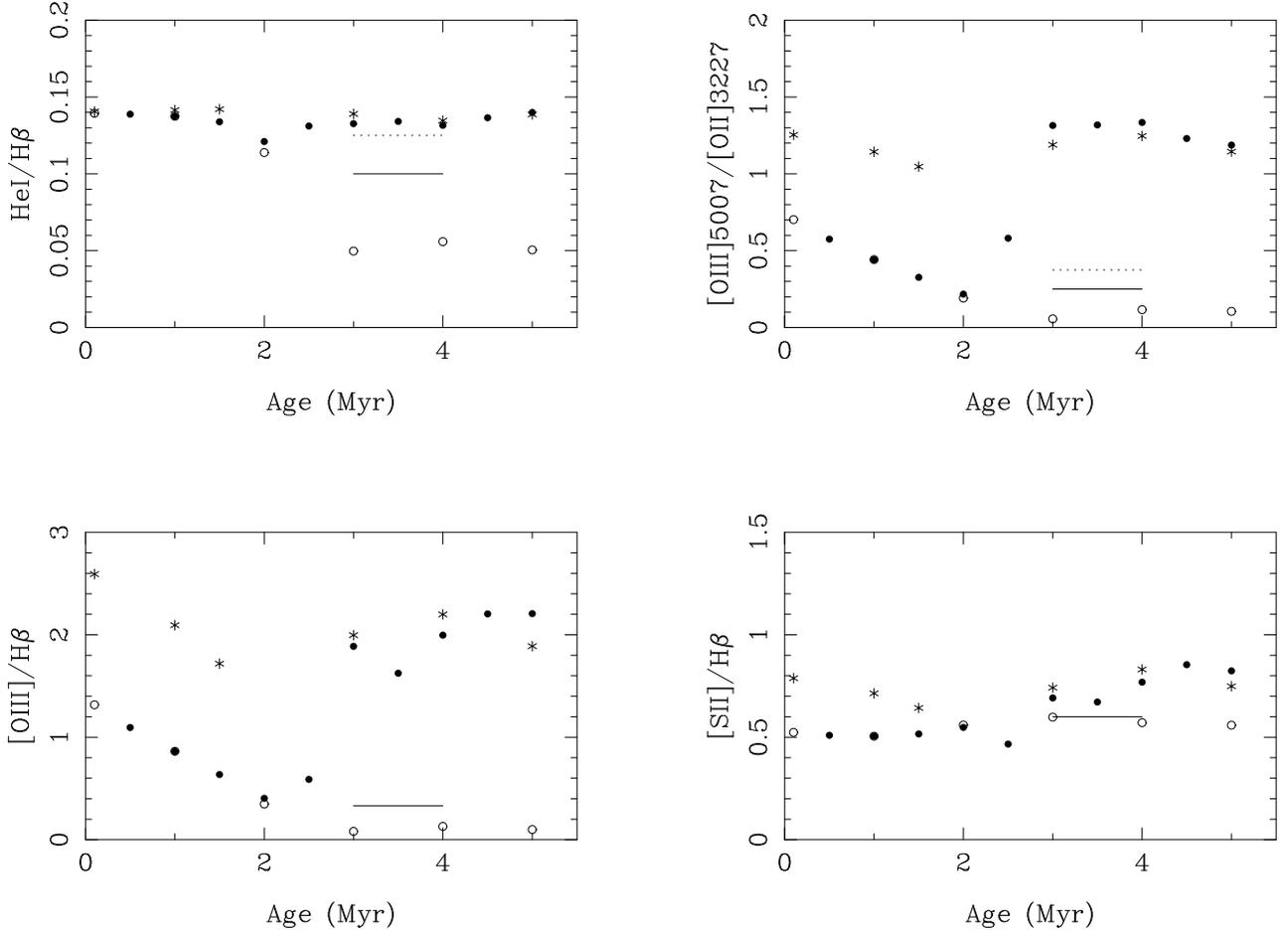}}

\figcaption{Comparison of the emission-line ratios obtained using ionizing radiation fields with 
different stellar atmosphere models. Predictions using the SED of SB99 with the
 Lejeune and Schmutz stellar atmospheres
are plotted with points, the SED of SB99 with Kurucz atmospheres with circles, and the
SED of Schaerer \& Vacca (1998) with 
stars.}

\end{figure}


\begin{figure} 

\centerline{\psfig{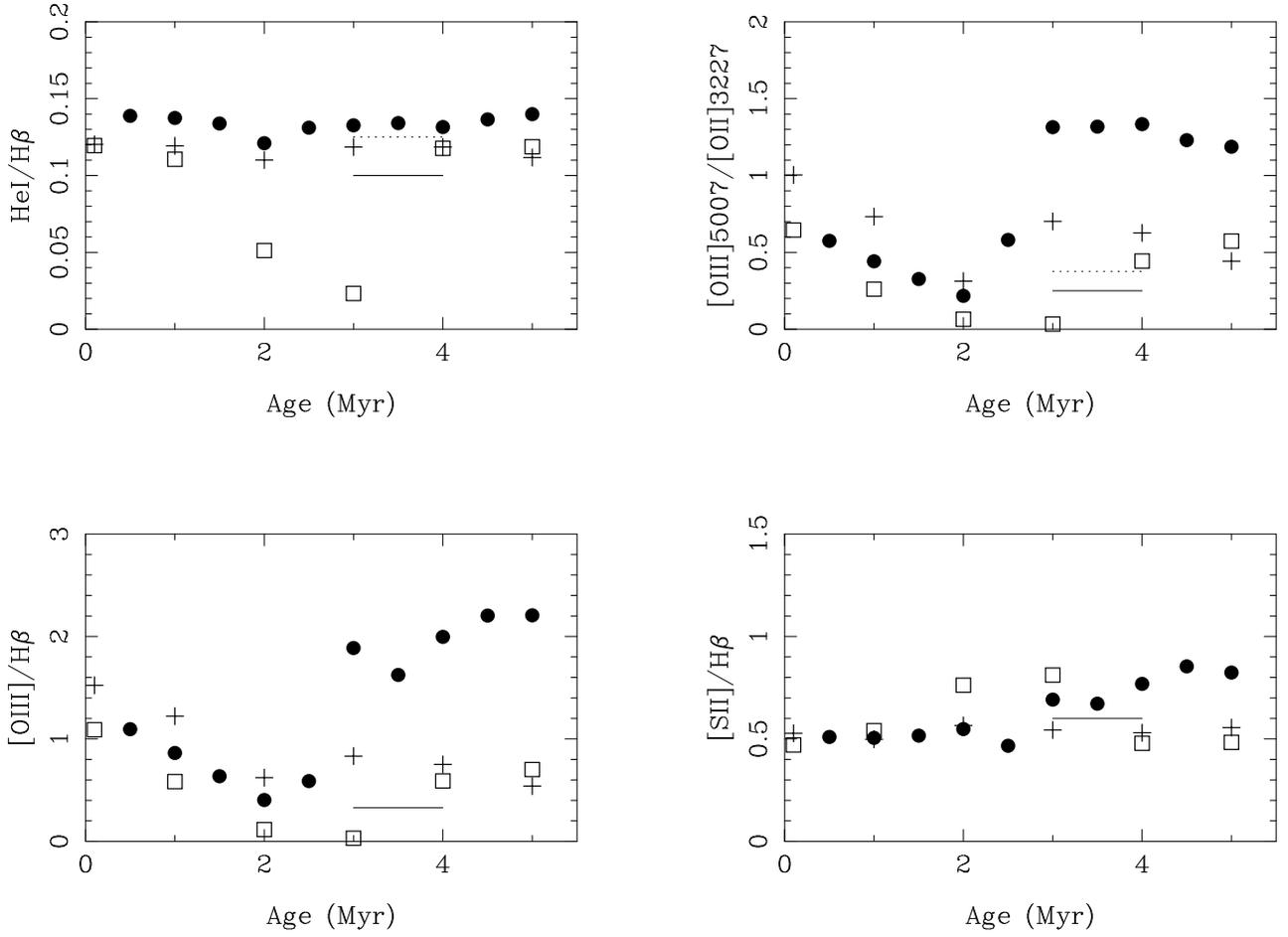}}

\figcaption{Comparison of the emission-line ratios obtained using ionizing radiation fields with 
different stellar atmosphere models. Predictions using the SED of SB99 with the
 Lejeune and Schmutz stellar atmospheres
are plotted with points, the SED of SB99 with Smith et al (2002) assuming that the stars 
have solar (twice solar) metallicity are plotted with crosses (squares).}

\end{figure}

\end{document}